# Optimal design of shell-graded-infill structures by a hybrid MMC-MMV approach

Chang Liu[1], Zongliang Du[1,2]*, Yichao Zhu[1], Weisheng Zhang[1], Xiaoyu Zhang[3], Xu Guo[1]*

[1]*State Key Laboratory of Structural Analysis for Industrial Equipment,*
*Department of Engineering Mechanics,*
*International Research Center for Computational Mechanics,*
*Dalian University of Technology, Dalian, 116023, P.R. China*

[2]*Mechanical & Aerospace Engineering,*
*University of Missouri, Columbia, MO 65211, United States of America*

[3]*Beijing Institute of Spacecraft System Engineering, CAST, Beijing, P.R. China*

## Abstract

In the present work, a hybrid MMC-MMV approach is developed for designing additive manufacturing-oriented shell-graded-infill structures. The key idea is to describe the geometry of a shell-graded-infill structure explicitly using some geometry parameters. To this end, a set of morphable voids is adopted to describe the boundary of the coating shell, while a set of morphable components combing with a coordinate perturbation technique are introduced to represent the graded infill distribution. Under such treatment, both the crisp boundary of the coating shell and the graded infill can be optimized simultaneously, with a small number of design variables. Numerical examples demonstrate the effectiveness of the proposed approach.

*Corresponding authors. E-mail: zldu@mail.dlut.edu.cn (Zongliang Du), guoxu@dlut.edu.cn (Xu Guo)

# 1. Introduction

Additive manufacturing (AM, also known as 3D printing) has an intrinsic ability to build structures with complex geometries. Topology optimization, on the other hand, provides a powerful tool to rationally and creatively distribute a certain amount of material within a prescribed design domain to achieve optimized structural performances. In order to fully leverage the capabilities of additive manufacturing technology, recent years witnessed an increasing interest to develop AM-oriented topology optimization approaches [1-7].

In the industrial 3D printing process (e.g., selective laser melting), to balance the fabrication cost and mechanical performance, the printed structures are designed to adopt a shell-infill form, i.e., a solid shell and a lattice infill constituting the exterior part (coating surface) and interior part of a structure, respectively. This kind of structure can effectively combine the advantages of the lightweight lattice and the efficient load-carrying capability of the coating structures simultaneously. Moreover, recent studies have shown that such structures also have excellent strength against buckling instability [8] resulting from unpredictable loading conditions and material defects or degradations [9]. For ease of design and manufacture, it was assumed in many previous works that the infill lattice is uniformly distributed. However, theoretical and experimental studies show that shell structures with *graded* lattice infills have many advantages compared with the ones composed of uniform infills. This fact has been demonstrated clearly in many nature-selected bio-materials. One of the promising examples, as shown in Fig. 1a, is that the strong shell and graded internal fibers of the bamboo greatly improve the bending stiffness [10]. More illustrations of “optimized shell-and-graded-infill designs” in nature can be found in bone [11] and tree trunk [12] presented in Fig. 1b and Fig. 1c respectively. Nowadays, topology optimization of shell structures with (graded) lattice infill has become a hot topic in the field of AM-oriented structural topology optimization.

In general, a shell-infill structure consists of a graded lattice infill and a solid shell exterior. For the lattice infill structures, Zhou and Li [13] developed a method to design graded microstructures with gradually varying effective properties, and applied this approach to design dental implants for bone remodeling [14]. Later on, Radman et al. [15] extended the above approach to the 3D cases as well as the design of functionally graded materials with prescribed variable thermal conductivity [16]. Wu et al. [17] presented a method to design bone-inspired non-uniform microstructures as porous infills by introducing an upper bound on a local volume measure under the Solid Isotropic Material with Penalization (SIMP) framework. Wang et al. [18] proposed a multi-scale concurrent design method, which is capable of optimizing the layout of the macroscale structure and graded micro-structures simultaneously, under the level set framework. Recently, Garner et al. [19] developed a new scheme to obtain well-connected graded microstructures with smoothly varying physical properties. For the shell structures (or coating structures), under the SIMP framework, Clausen et al. [20] proposed a method to design coating structures composed of a solid shell and a uniform base material. Wadbro and Niu [21] developed a multiscale method which can concurrently design the solid coating structures and periodic infill patterns. Luo et al. [11] adopted an erosion-based interface identification method to design shell–infill structures and further improved the existing two-step filtering/projection process. Under the level-set framework, Wang et al. [22] presented a topology optimization method for the design of coating structures. Fu et al. [23] proposed a novel topology optimization formulation for shell-infill structures based on a distance regularized parametric level-set method. The above works, however, mainly focus on the design of coating structures, while the infill structures are assumed to be composed of weak solid materials or theoretically infinite-small periodic lattice cell without definite length scale, which obviously cannot be used for additive manufacturing directly.

In order to resolve the aforementioned issues, recently, some advanced approaches that can simultaneously design the shell and spatially graded infill structures have been developed. Wu et al. [24] proposed a complete solution to the optimal design of shell–infill structures by introducing two design fields for concurrently evolving the shell interface and non-uniform infill in density-based topology optimization. This method can obtain infill structures following the principal stress directions and is remarkably robust to parameter and geometric variations. However, it requires successive filter operations to obtain the shell interface and the global material volume constraint cannot be strictly satisfied. Later on, Groen et al. [25] developed a homogenization-based approach to design coated structures with orthotropic infills, which allows the modeling of designs with complex microstructures on a relatively coarse mesh. Remarkably, near-optimal designs can be obtained while the computational cost is reduced by at least one order of magnitude compared to the density-based shell–infill structure design method. Nevertheless, this method is executed through successive processes (e.g., homogenization, double smoothing and projection) and the origin of graded infill is fixed as a square unit-cell with a rectangular hole. In addition, Wang et al. [26] developed a level set based shell–infill structures design method, in which the spatially-varying graded mesostructures are obtained through the variations of the sizes of the structural components in a prototype unit cell. In this manner, although the connectivity of self-similar graded mesostructures is achieved to some extent, mismatches inevitably exist at the interface of adjacent unit cells. Therefore, for real-world applications, besides tedious post-processing required to render the smoothness of the optimized designs, the self-similar property of the geometry of infill may also restrict the design space. In summary, although impressive optimized shell-infill designs have been obtained by these approaches, there still remains room for further improvements, such as simplifying the construction procedure of shell-graded-infill structures and generating smoothly graded infills.

From the authors' point of view, the aforementioned problems can be solved from a more geometrical perspective. The key point is to describe the geometry of both the graded lattice structure and the coating structure in a parameter-based explicit way. To be specific, as shown in Fig. 2, following the idea of continuum mechanics, a non-uniform lattice structure $\Omega^{\mathrm{g}}(\boldsymbol{x})$ can be regarded as the current configuration of some uniform lattice (i.e., the initial configuration $\Omega^{\mathrm{p}}(\boldsymbol{x})$) with respect to a specific deformation gradient field $\boldsymbol{F}(\boldsymbol{x})$. On the other hand, by introducing the coordinate transformation field $\boldsymbol{F}(\boldsymbol{x})$ in the current configuration, i.e., $\widetilde{\boldsymbol{x}}=\boldsymbol{F}(\boldsymbol{x})$, the non-uniform lattice can be expressed as $\Omega^{\mathrm{p}}(\widetilde{\boldsymbol{x}})$, which is in the same form as the initial uniform configuration. This inspires us that the non-uniform lattice structure can be generated by a uniform lattice combined with a non-uniform coordinate transformation field. Expanding the transformation field by some basis functions, a smooth optimized graded infill can be determined through optimizing the layout of a parent uniform lattice and the coefficients of basis functions [27]. Besides, from a geometrical point of view, the coating shell structure can be constructed by a "difference" operation between two similar structures. As shown in Fig. 3, for a given structure $\Omega$, moving its boundary points inward by a distance $\Delta d$, a similar structure $\Omega^{\mathrm{ext}}$ can be obtained. It is obvious that the difference set of these two structures $\Omega\backslash\Omega^{\mathrm{ext}} \triangleq \Omega^{\mathrm{shell}}$ is nothing but a coating shell structure with thickness $\Delta d$. As will be shown in the rest parts of this work, by combining this construction procedure with the so-called Moving Morphable Components (MMC) [28] and Moving Morphable Voids (MMV) methods [29], an explicit geometric description of a shell-graded-infill structure can be achieved, and this treatment may provide great potential for solving the aforementioned challenging issues associated with the integrated optimization of shell-graded-infill structures.

The Moving Morphable Components (MMC) [28] and Moving Morphable Voids (MMV) methods [29], which describe the layout of a structure in an explicit way, are considered to be promising approaches for solving topology optimization problems where various geometry features are concerned [11]. Actually, in the MMC and MMV approach, the structural topologies are represented by a set of explicit geometric parameters (e.g., the lengths, widths, oblique angles of the components), and the advantages of these approaches have been demonstrated in the design of AM-oriented self-supporting structures [5] and graded structures [27], as well as realizing length scale/thickness control [30, 31] in topology optimization problems, just to name a few. Nevertheless, in order to apply the MMC/MMV method for designing AM-oriented shell-infill structures, the following issues still need to be resolved: (1) How to construct the topology description function of a shell-graded-infill structure by combining the MMC and MMV methods? (2) How to formulate the optimization problem to concurrently design the shell and infill in terms of related geometric parameters? (3) For specific problems, e.g., minimum compliance design, how to perform the sensitivities analysis efficiently when both MMC and MMV based geometry description is employed?

To address the afore-mentioned issues, in this work, a hybrid MMC-MMV method is developed for designing AM-oriented shell-graded-infill structures. To be specific, the solid coating shell is constructed with the help of MMVs while the graded lattice infills are described based on MMCs. In this way, the shell-graded-infill structures can be explicitly described by only a set of parameters describing the geometries of the components and voids in the MMC and MMV formulations. Moreover, both the exterior solid shell and the interior graded lattice infill can be optimized simultaneously with the use of the analytical sensitivity information. Comparisons between the optimized shell-graded-infill designs and their counterparts with uniform infills are made to illustrate the effectiveness of the proposed approach.

The rest of the paper is organized as follows. In Section 2, the hybrid MMC-MMV method for describing shell-graded-infill structures is presented. In Section 3, the mathematical formulation of the considered optimization problem is presented. A discussion on finite element (FE) discretization and sensitivity analysis is then provided in Section 4. In Section 5, the effectiveness of the proposed approach is verified through three representative numerical examples. Finally, some concluding remarks are provided.

## 2. Description of shell-graded-infill structures by a hybrid MMC-MMV method

### *2.1 MMC/MMV-based topology optimization methods*

The Moving Morphable Components (MMC) approach and the Moving Morphable Voids (MMV) approach for structural topology optimization were proposed in [28, 29] and further extended in [30, 32-41]. In the MMC/MMV approaches, a number of morphable structural components/voids are adopted as building blocks of the topological design, the optimized structural topology can be determined by optimizing the explicit geometry parameters characterizing the sizes, shapes, and layout of the introduced components/voids (see Fig. 4 and Fig. 5 for reference) in a unified framework. Compared with traditional pixel [42, 43] or node point-based [44, 45] solution approaches, MMC/MMV methods usually involve a relatively small number of design variables and have the capability of explicitly controlling the geometry features (e.g., the inclined angle of a component, the pattern of a infill design and the thickness of a coating shell) of a structure. For the sake of completeness, the basic idea and numerical implementation of the MMC/MMV method will be briefly sketched in the following subsection. The readers are referred to [28-30, 32-38] for more technical details. It should be noted that, the MMC/MMV methods can be implemented in both Eulerian description-based framework [28, 30, 33, 35-37] and Lagrangian description-based framework [29, 34]. In the present study, both structural components and voids are described using a unified Eulerian mesh.

Under the *Eulerian description-based topology optimization framework*, the region occupied by a given amount of solid material can be described in terms of a so-called topology description function (TDF) $\phi^{\mathrm{s}}$ as follows:

$$\begin{cases} \phi^{\mathrm{s}}(\boldsymbol{x}) > 0, \text{if } \boldsymbol{x} \in \Omega^{\mathrm{s}}, \\ \phi^{\mathrm{s}}(\boldsymbol{x}) = 0, \text{if } \boldsymbol{x} \in \partial\Omega^{\mathrm{s}}, \\ \phi^{\mathrm{s}}(\boldsymbol{x}) < 0, \text{if } \boldsymbol{x} \in \mathrm{D}\backslash(\Omega^{\mathrm{s}} \cup \partial\Omega^{\mathrm{s}}), \end{cases} \tag{2.1}$$

where D represents a prescribed design domain and $\Omega^{\mathrm{s}} \subset \mathrm{D}$ is a subset of D occupied by the solid material.

As illustrated by Fig. 4c, in the MMC approach, it is assumed that $\Omega^{\mathrm{s}}$ is comprised of $nc$ components made of solid material. Under this circumstance, it yields that $\phi^{\mathrm{s}}(\boldsymbol{x}) = \max\left(\phi_1^{\mathrm{C}}(\boldsymbol{x}), \cdots, \phi_{nc}^{\mathrm{C}}(\boldsymbol{x})\right)$ with $\phi_i^{\mathrm{C}}(\boldsymbol{x})$ denoting the TDF of the $i$-th component [28] as:

$$\begin{cases} \phi_i^{\mathrm{C}}(\boldsymbol{x}) > 0, \text{if } \boldsymbol{x} \in \Omega_i^{\mathrm{s}}, \\ \phi_i^{\mathrm{C}}(\boldsymbol{x}) = 0, \text{if } \boldsymbol{x} \in \partial\Omega_i^{\mathrm{s}}, \\ \phi_i^{\mathrm{C}}(\boldsymbol{x}) < 0, \text{if } \boldsymbol{x} \in \mathrm{D}\backslash(\Omega_i^{\mathrm{s}} \cup \partial\Omega_i^{\mathrm{s}}), \end{cases} \tag{2.2}$$

where $\Omega_i^{\mathrm{s}} \subset \mathrm{D}$ is the region occupied by the $i$-th component and obviously $\Omega^{\mathrm{s}} = \cup_{i=1}^{nc} \Omega_i^{\mathrm{s}}$.

For the two-dimensional (2D) case (the case considered in the present work), the function $\phi_i^{\mathrm{C}}(\boldsymbol{x})$ can be adopted as:

$$\phi_i^{\mathrm{C}}(x, y) = 1 - \left(\frac{x'}{a_i}\right)^p - \left(\frac{y'}{b_i(x')}\right)^p \tag{2.3}$$

with

$$\begin{Bmatrix} x' \\ y' \end{Bmatrix} = \begin{bmatrix} \cos\theta_i & \sin\theta_i \\ -\sin\theta_i & \cos\theta_i \end{bmatrix} \begin{Bmatrix} x - x_{0i} \\ y - y_{0i} \end{Bmatrix} \tag{2.4}$$

and $p$ is a relatively large even integer (e.g. $p = 6$). In Eq. (2.3) and Eq. (2.4), the symbols $(x_{0i}, y_{0i})$, $a_i$, $b_i(x')$ and $\theta_i$ denote the coordinates of the center, the half-length, the variable half-width and the inclined angle (measured from the horizontal axis anti-clockwisely) of the $i$-th component (see Fig. 6 for reference), respectively. It should also be noted that the variation of the width $b_i(x')$ is measured with respect to a local coordinate system and $b_i(x')$ can take different forms [33]. In the present work, it is chosen as

$$b_i(x') = \frac{t_i^1 + t_i^2}{2} + \frac{t_i^2 - t_i^1}{2a_i} x', \tag{2.5}$$

where $t_i^1$ and $t_i^2$ are parameters used to describe the thicknesses of the $i$-th component.

Under the above geometry representation scheme, the layout of a structure in the MMC method can be solely determined by a design variable vector $\boldsymbol{D} = \left((\boldsymbol{D}_1^{\mathrm{C}})^{\top}, \ldots, (\boldsymbol{D}_i^{\mathrm{C}})^{\top}, \ldots (\boldsymbol{D}_{nc}^{\mathrm{C}})^{\top}\right)^{\top}$, where $\boldsymbol{D}_i^{\mathrm{C}} = \left(x_{0i}^{\mathrm{C}}, y_{0i}^{\mathrm{C}}, a_i, \boldsymbol{d}_i^{\top}, \theta_i\right)^{\top}$ and the symbol $\boldsymbol{d}_i = (t_i^1, t_i^2)^{\top}$ denotes the vector of design parameters associated with $b_i(x')$.

In the MMV approach, it holds that $\Omega^{\mathrm{s}} = \mathrm{D} \backslash \cup_{j=1}^{nv} \Omega_j^{\mathrm{V}}$, where $\Omega_j^{\mathrm{V}}$ is $j$-th void region enclosed by a set of closed parametric B-spline curves $S_j$, $j = 1, \ldots, nv$, respectively. The geometric parameters related to the curve $S_j$ include the coordinates of the central point $\boldsymbol{P}_{0j} = (x_{0j}^{\mathrm{V}}, y_{0j}^{\mathrm{V}})$ and the coordinates of the B-spline control points $\boldsymbol{P}_j^k = \left(P_{jx}^k, P_{jy}^k\right)^{\top}$ and $\boldsymbol{P}_0^k = \boldsymbol{P}_n^k$, $k = 0, \ldots, n$ (in this study, $n = 12$). As shown in

Fig. 7, the control points $\boldsymbol{P}_j^k$ are uniformly distributed along the circumference following:

$$P_{jx}^k = x_{0j}^{\mathrm{V}} + d_j^k \cos(\psi^k), \tag{2.6a}$$

$$P_{jy}^k = y_{0j}^{\mathrm{V}} + d_j^k \sin(\psi^k), \tag{2.6b}$$

where

$$\psi^k = \frac{2k\pi}{n}, k = 0, \ldots, n, \tag{2.7}$$

is the pre-specified inclined angle (measured from the horizontal axis anti-clockwisely) of radial $\overrightarrow{\boldsymbol{P}_{0j}\boldsymbol{P}_j^k}$ with respect to radiation $\overrightarrow{\boldsymbol{P}_{0j}\boldsymbol{P}_j^0}$. In Eq. (2.6), $d_j^k$ is the distance from $\boldsymbol{P}_j^k$ to $\boldsymbol{P}_{0j}$, which is to be optimized during the optimization process.

For a boundary point $\boldsymbol{P}_{\mathrm{b}j} \in S_j$, the distance from $\boldsymbol{P}_{\mathrm{b}j}$ to $\boldsymbol{P}_{0j}$ can be interpolated as

$$r_j(\psi) = \left\|\overrightarrow{\boldsymbol{P}_{0j}\boldsymbol{P}_{\mathrm{b}j}(\mu(\psi))}\right\| = \left\|\sum_{k=0}^{n} N_{k,q}\big(\mu(\psi)\big)\boldsymbol{P}_j^k\right\|, \tag{2.8}$$

where $\psi$ denotes the inclined angle of the radial $\overrightarrow{\boldsymbol{P}_{0j}\boldsymbol{P}_{\mathrm{b}j}}$ illustrated by Fig. 7. As shown in [29, 35, 37], in Eq. (2.8), $N_{k,q}(\mu)$ is the $q$-order ($q = 2$ in this study) B-spline basis function with $\mu \in [0,1]$.

Under the Euler description framework, the TDF of $j$-th void in the MMV method can be expressed as:

$$\phi_j^{\mathrm{V}}(x, y) = \sqrt{\big(x - x_{0j}^{\mathrm{V}}\big)^2 + \big(y - y_{0j}^{\mathrm{V}}\big)^2} - r_j(\psi). \tag{2.9}$$

Accordingly, the domain $\Omega_j^{\mathrm{V}}$ occupied by the $j$-th star-shaped void can be described in terms of $\phi_j^{\mathrm{V}}, j = 1, \ldots, nv$ as:

$$\begin{cases} \phi_j^{\mathrm{V}}(\boldsymbol{x}) < 0, \text{if } \boldsymbol{x} \in \Omega_j^{\mathrm{V}}, \\ \phi_j^{\mathrm{V}}(\boldsymbol{x}) = 0, \text{if } \boldsymbol{x} \in \partial\Omega_j^{\mathrm{V}}, \\ \phi_j^{\mathrm{V}}(\boldsymbol{x}) > 0, \text{if } \boldsymbol{x} \in \mathrm{D}\backslash(\Omega_j^{\mathrm{V}} \cup \partial\Omega_j^{\mathrm{V}}). \end{cases} \tag{2.10}$$

Therefore, the TDF of the whole structure in the MMV approach can be calculated as $\phi^{\mathrm{s}}(\boldsymbol{x}) = \min\Big(\phi_1^{\mathrm{V}}(\boldsymbol{x}), \cdots, \phi_{nv}^{\mathrm{V}}(\boldsymbol{x})\Big)$.

In the MMV-based approach, the layout of a structure can be solely determined by a design variable vector $\boldsymbol{D} = \left((\boldsymbol{D}_1^{\mathrm{V}})^{\top}, \ldots, (\boldsymbol{D}_j^{\mathrm{V}})^{\top}, \ldots (\boldsymbol{D}_{nv}^{\mathrm{V}})^{\top}\right)^{\top}$, where $\boldsymbol{D}_j^{\mathrm{V}} = \left(x_{0j}^{\mathrm{V}}, y_{0j}^{\mathrm{V}}, \boldsymbol{r}_j^{\top}\right)^{\top}$ with $\boldsymbol{r}_j = \left(d_j^1, \ldots, d_j^n\right)^{\top}$. It is worth noting that the TDFs of more complicated-shaped components and voids can also be constructed in a systematic way. We refer the readers to [32, 35] for more details on this aspect.

### *2.2 Explicit geometry description of shell-graded-infill structures*

In this subsection, the TDF characterizing the geometry of shell-graded-infill structures is described by a hybrid MMC-MMV method. In particular, the MMC description is used to obtain a graded lattice background while the MMV description is used to determine the coating shell as well as the boundary of the structure. The geometry description of the integrated shell-graded-infill structure can be achieved by taking some operations on the respective TDFs associated with each component and void.

To introduce a specific size to meet the requirement of manufacturing, without loss of generality, the graded lattice background is assumed to be in a rectangle-shape and generated by a uniform lattice structure incorporating a coordinate transformation filed. The uniform lattice is assumed to be a $n_x \times n_y$ array of a prototype unit cell composed of $nc$ components, as shown in Fig. 8. The TDF of the $k$-th ($k = 1, \ldots, nc$) component in the $l$-th ($l = 1, \ldots, n_x \times n_y \triangleq ncell$) unit cell is expressed as

$$\phi^{k,l}(x, y) = 1 - \sqrt[p]{\left(\frac{x'}{a_k}\right)^p + \left(\frac{y'}{b_k(x')}\right)^p}, \tag{2.11}$$

with

$$\begin{Bmatrix} x' \\ y' \end{Bmatrix} = \begin{bmatrix} \cos\theta_k & \sin\theta_k \\ -\sin\theta_k & \cos\theta_k \end{bmatrix} \begin{Bmatrix} x - x^{k,l} \\ y - y^{k,l} \end{Bmatrix}, \tag{2.12}$$

where $(x^{k,l}, y^{k,l})^{\top}$ is the coordinates of the centeral point and can be easily obtained based on $(x^{k,1}, y^{k,1})^{\top}$. It is worth noting that to facilitate the use of MMC and MMV approaches within a unified framework, a $p$ −th root operation is introduced in Eq. (2.11) to make the TDF of the components changing more gently.

It is worthwhile to mention that, different from the treatment in previous work [27],

in order to improve the deformability of each component, in the present work, instead of using a global coordinate transformation field, we adopt an independent coordinate transformation mapping for each component in the prototype unit cell. To be specific, for the $k$-th component, we have:

$$\begin{pmatrix} \tilde{x}^k \\ \tilde{y}^k \end{pmatrix} = \begin{pmatrix} x + f^k(x,y) \\ y + g^k(x,y) \end{pmatrix}, \tag{2.13}$$

where the symbols $f^k(x,y)$, $g^k(x,y)$ are the corresponding coordinate perturbation functions. In the present paper, it is assumed that $f^k(x,y) = f^k(x)$, $g^k(x,y) = g^k(y)$ for simplicity and only trigonometric basis functions are adopted in numerical implementation. In particular, the corresponding coordinate perturbation functions (CPFs) are expressed as

$$\begin{cases} f^k(x) = \sum_{r=1}^{n_1} \alpha_{r,1}^k \cos\left(\frac{(r-1)\pi}{L}\left(x - \frac{L}{2}\right)\right) + \alpha_{r,2}^k \sin\left(\frac{(r-1)\pi}{L}\left(x - \frac{L}{2}\right)\right), \\ g^k(y) = \sum_{t=1}^{n_2} \beta_{t,1}^k \cos\left(\frac{(t-1)\pi}{H}\left(y - \frac{H}{2}\right)\right) + \beta_{t,2}^k \sin\left(\frac{(t-1)\pi}{H}\left(y - \frac{H}{2}\right)\right). \end{cases} \tag{2.14}$$

In Eq. (2.14), $L$ and $H$ are the characteristic length and width of the design domain, $\alpha_{r,1}^k, \alpha_{r,2}^k, \beta_{t,1}^k, \beta_{t,2}^k$ are the corresponding coefficients to be determined through optimization.

In this manner, the graded infill background can be described by a vector $\boldsymbol{D}^{\mathrm{infill}} = \left(\left(\overline{\boldsymbol{D}}_1^{\mathrm{C}}\right)^\top, \ldots, \left(\overline{\boldsymbol{D}}_k^{\mathrm{C}}\right)^\top, \ldots (\overline{\boldsymbol{D}}_{nc}^{\mathrm{C}})^\top\right)^\top$, where the symbol $\overline{\boldsymbol{D}}_k^{\mathrm{C}} = (x^{k,1}, y^{k,1}, a_k, \theta_k, t_k^1, t_k^2, (\boldsymbol{\alpha}^k)^\top, (\boldsymbol{\beta}^k)^\top)^\top$ contains the parameters associated with the $k$-th component in the prototype unit cell (i.e., unit cell (1,1)) and $\boldsymbol{\alpha}^k = \left(\alpha_{1,1}^k, \alpha_{1,2}^k, \ldots, \alpha_{r,1}^k, \alpha_{r,2}^k, \ldots, \alpha_{n_1,1}^k, \alpha_{n_1,2}^k\right)^\top$, $\boldsymbol{\beta}^k = \left(\beta_{1,1}^k, \beta_{1,2}^k, \ldots, \beta_{t,1}^k, \beta_{t,2}^k, \ldots, \beta_{n_2,1}^k, \beta_{n_2,2}^k\right)^\top$ are the vectors of the coefficients associated with the perturbation basis functions along the two coordinate directions. Furthermore, the TDFs of the graded lattice background (see Fig. 9 for reference) can be expressed as

$$\phi^{\mathrm{g-s}}\left(\boldsymbol{x}; \boldsymbol{D}^{\mathrm{infill}}\right) = \max_{k=1}^{nc}\left(\phi^k\left(\tilde{\boldsymbol{x}}^k; \overline{\boldsymbol{D}}_k^{\mathrm{C}}\right)\right) = \max_{k=1}^{nc}\max_{l=1}^{ncell}\left(\phi^{k,l}\left(\tilde{\boldsymbol{x}}^k, \overline{\boldsymbol{D}}_k^{\mathrm{C}}\right)\right). \tag{2.15}$$

Once $\phi^{\mathrm{g-s}}$ is obtained, the next step is to construct the TDFs of the shell structure and the shell-graded-infill structure. Due to the explicit description of the structural boundary, it is easy to construct the TDF of the shell structure based on the MMV method. In particular, as shown in Fig. 10a, first we can generate the TDF of a solid region (i.e., $\Omega_0$ in Fig. 10a) using the MMV-based description in the following way:

$$\phi_0 = \min\left(\phi_1^{\mathrm{V}}(\boldsymbol{x}), \cdots, \phi_j^{\mathrm{V}}(\boldsymbol{x}), \cdots, \phi_{nv}^{\mathrm{V}}(\boldsymbol{x})\right), \tag{2.16}$$

where $nv$ is the number of the initial voids. The symbol $\phi_j^{\mathrm{V}}(\boldsymbol{x})$ is the TDF of the $j$-th void ($j = 1, \dots, nv$), which can be calculated by Eq. (2.9). The design variable vector of the $j$-th void is $\boldsymbol{D}_j^{\mathrm{V}} = \left(x_{0j}^{\mathrm{V}}, y_{0j}^{\mathrm{V}}, \boldsymbol{r}_j^{\mathrm{T}}\right)^{\mathrm{T}}$. Then we can expand the $j$-th void by perturbing the distance vector $\boldsymbol{r}_j = \left(d_j^1, \dots, d_j{}^n\right)^{\mathrm{T}}$ as $\boldsymbol{r}_j^{\mathrm{ext}} = \left(d_j{}^1 + \Delta d, \dots, d_j{}^n + \Delta d\right)^{\mathrm{T}}, j = 1, \dots, nv$. This treatment yields a new structure $\Omega_0^{\mathrm{ext}}$, as illustrated in Fig. 10b, with its TDF defined as

$$\phi_0^{\mathrm{ext}} = \min\left(\phi_1^{\mathrm{V-ext}}(\boldsymbol{x}), \cdots, \phi_j^{\mathrm{V-ext}}(\boldsymbol{x}), \cdots, \phi_{nv}^{\mathrm{V-ext}}(\boldsymbol{x})\right), \tag{2.17}$$

where the design variable vector of the $j$-th void in this intermediate structure is $\boldsymbol{D}_j^{\mathrm{V-ext}} = \left(x_{0j}^{\mathrm{V}}, y_{0j}^{\mathrm{V}}, \left(\boldsymbol{r}_j^{\mathrm{ext}}\right)^{\mathrm{T}}\right)^{\mathrm{T}}$. Furthermore, the region occupied by the coating shell can be constructed as $\Omega^{\mathrm{shell}} = \Omega_0 \backslash \Omega_0^{\mathrm{ext}}$ as shown in Fig. 10c.

Although we can easily obtain the graded lattice structure $\Omega^{\mathrm{g-s}}$ and the shell structure $\Omega^{\mathrm{shell}}$ separately, the essential point is to combine them to form a shell-graded-infill structure under a unified framework. Actually, this cannot be achieved by a direct Boolean operation on $\Omega^{\mathrm{g-s}}$ and $\Omega^{\mathrm{shell}}$. For the graded lattice infill, only the portion between the inner boundaries of the shell structure (the area between the dotted lines in Fig. 10c) is useful for the construction of the shell-graded-infill structure. In principle, this requires an accurate identification of the internal (the dotted lines in Fig. 10c) and external boundaries (the solid lines in Fig. 10c) of the shell structure. Therefore, we adopt the following step-by-step procedures to distinguish the inner and outer boundaries of the shell structure and construct the shell-graded-infill structures.

First, we can obtain the dual structure of $\Omega_0^{\mathrm{ext}}$ as $\overline{\Omega_0^{\mathrm{ext}}} = \mathrm{D} \backslash \Omega_0^{\mathrm{ext}}$ and its TDF can be written as $-\phi_0^{\mathrm{ext}}$. From Fig. 10d, it is obvious that the boundary of the void regions inside $\overline{\Omega_0^{\mathrm{ext}}}$ is exactly the same as the inner boundary of the shell structure shown in Fig. 10c. Meanwhile, the structural boundary of $\Omega_0$ in Fig. 10a coincides with the outer boundary of the shell structure in Fig. 10c.

Next, we can combine $\overline{\Omega_0^{\mathrm{ext}}}$ (Fig. 10d) and $\Omega^{\mathrm{g-s}}$ (Fig. 9b) to obtain a new region $\Omega_1 = \overline{\Omega_0^{\mathrm{ext}}} \cup \Omega^{\mathrm{g-s}}$. As shown in Fig. 10e, in $\Omega_1$, the region between the inner boundaries of the shell structure $\Omega^{\mathrm{shell}}$ is filled with the graded lattice structure, while the rest part in the design domain is solid. Accordingly, the TDF of $\Omega_1$ can be defined as

$$\phi_1 = \max(-\phi_0^{\mathrm{ext}}, \phi^{\mathrm{g-s}}). \tag{2.18}$$

Finally, the shell-graded-infill structure shown in Fig. 10f can be obtained by removing the redundant part with respect to $\Omega_0$. This can be simply achieved by realizing the fact that $\Omega^{\mathrm{s}} = \Omega_1 \cap \Omega_0$, and the TDF of the shell-graded-infill structure can be constructed as

$$\phi^{\mathrm{s}} = \min(\phi_0, \phi_1). \tag{2.19}$$

Although the above process seems to be complicated, it is actually a clear fact that $\Omega^{\mathrm{s}} = \Omega_0 \cap (\Omega_{\mathrm{D}} \backslash \Omega_0^{\mathrm{ext}} \cup \Omega^{\mathrm{g-s}})$, and correspondingly the TDF of the shell-graded-infill structure reads

$$\phi^{\mathrm{s}} = \min\left(\phi_0, \max\left(-\phi_0^{\mathrm{ext}}, \phi^{\mathrm{g-s}}\right)\right). \tag{2.20}$$

Notably, the above operations are purely based on the structural geometry, and all design variables have exact geometric meanings. During the construction process of the shell (coating) structure, only a parameter $\Delta d$ associated with the shell-thickness is introduced. Compared with the treatment in previous works [20-24], the proposed approach can achieve the same goal in a more straightforward and simple way. It is also worth noting that, although the value of $\Delta d$ is fixed in the present work, it can also be optimized to obtain an optimal thickness of the coating shell in the proposed problem formulation.

In summary, in the hybrid MMC-MMV approach, a shell-graded-infill structure can be explicitly described by a design variable vector $\boldsymbol{D} = \left(\left(\boldsymbol{D}^{\mathrm{infill}}\right)^{\top}, \left(\boldsymbol{D}^{\mathrm{shell}}\right)^{\top}\right)^{\top}$

with $\boldsymbol{D}^{\text{infill}} = \left((\bar{\boldsymbol{D}}_1^{\text{C}})^\top, \dots, (\bar{\boldsymbol{D}}_k^{\text{C}})^\top, \dots (\bar{\boldsymbol{D}}_{nc}^{\text{C}})^\top\right)^\top$ and $\boldsymbol{D}^{\text{shell}} = \left((\boldsymbol{D}_1^{\text{V}})^\top, \dots, (\boldsymbol{D}_j^{\text{V}})^\top, \dots (\boldsymbol{D}_{nv}^{\text{V}})^\top\right)^\top$, respectively.

## 3. Problem formulation

In the present study, the shell-graded-infill structures are designed to minimize the structural compliance under the volume constraint of available solid material. Besides, an additional volume constraint of the infill region was introduced to prevent the optimized shell-infill structures from degenerating to solid structures, as will be demonstrated in subsequent numerical examples. Under this circumstance, the corresponding problem formulation can be written as:

$$\begin{aligned}
&\text{Find } \boldsymbol{D} = \left((\boldsymbol{D}^{\text{infill}})^\top, (\boldsymbol{D}^{\text{shell}})^\top\right)^\top, \boldsymbol{u}(\boldsymbol{x}) \in \boldsymbol{\mathcal{H}}^1(D) \\
&\text{Minimize} \quad C = \int_D H\big(\phi^{\text{s}}(\boldsymbol{x};\boldsymbol{D})\big)\boldsymbol{f}\cdot\boldsymbol{u}\,\text{dV} + \int_{\Gamma_{\text{t}}} \boldsymbol{t}\cdot\boldsymbol{u}\,\text{dS} \\
&\text{S.t.} \\
&\int_D \mathbb{E}(H(\phi^{\text{s}}(\boldsymbol{x};\boldsymbol{D}))):\boldsymbol{\varepsilon}(\boldsymbol{u}):\boldsymbol{\varepsilon}(\boldsymbol{v})\,\text{dV} = \int_D H\big(\phi^{\text{s}}(\boldsymbol{x};\boldsymbol{D})\big)\boldsymbol{f}\cdot\boldsymbol{v}\,\text{dV} \\
&\qquad + \int_{\Gamma_{\text{t}}} \boldsymbol{t}\cdot\boldsymbol{v}\,\text{dS}, \quad \forall \boldsymbol{v} \in \mathcal{U}_{\text{ad}}, \\
&V = \int_D H\big(\phi^{\text{s}}(\boldsymbol{x};\boldsymbol{D})\big)\,\text{dV} \le \bar{V}, \\
&V_{\text{in}} = \int_D H\left(\phi_0^{\text{ext}}(\boldsymbol{x};\boldsymbol{D}^{\text{shell}})\right)\text{dV} \ge \underline{V}, \\
&\boldsymbol{D} \subset \mathcal{U}_{\boldsymbol{D}}, \\
&\boldsymbol{u} = \bar{\boldsymbol{u}}, \text{on } \Gamma_{\text{u}},
\end{aligned} \tag{3.1}$$

where $D$, $\boldsymbol{f}$, $\boldsymbol{t}$ and $\bar{\boldsymbol{u}}$ are the design domain, the body force density, the prescribed surface traction on Neumann boundary $\Gamma_{\text{t}}$ and the prescribed displacement on Dirichlet boundary $\Gamma_{\text{u}}$, respectively. In Eq. (3.1), $\boldsymbol{\mathcal{H}}^1(D)$ denotes the first-order Sobolev space and $H = H(x)$ is the Heaviside function with $H = 1$ if $x > 0$, otherwise $H = 0$. Besides, $\mathbb{E} = \mathbb{E}(H(\phi^{\text{s}}(\boldsymbol{x};\boldsymbol{D})))$ is the fourth-order elasticity tensor of the material locating at point $\boldsymbol{x}$ and $\mathcal{U}_{\mathrm{ad}} = \{\boldsymbol{v} | \boldsymbol{v} \in \boldsymbol{H}^1(D), \boldsymbol{v} = \boldsymbol{0} \text{ on } \mathrm{S_u}\}$ is the admissible set of virtual displacement vector $\boldsymbol{v}$.

In Eq. (3.1), the second constraint is the well-known material usage constraint, with $V$ and $\bar{V}$ denoting the actual material usage and its upper limit. The third constraint is the volume constraint of the infill region, which requires that the volume of the region occupied by the infill structure should be greater than a prescribed lower bound $\underline{V}$. The symbol $\phi_0^{\mathrm{ext}}$ is the TDF of the structure corresponding to Fig. 10b and the volume of the infill region is denoted as $V_{\mathrm{in}}$. It should be noted that both the material usage and the volume of the infill region can be controlled during the optimization process, which is also an advantage of the proposed method.

In this manner, both the coating shell structure and the graded infill can be concurrently optimized by solving Eq. (3.1) in terms of geometric parameters.

## 4. Finite element (FE) discretization and sensitivity analysis

In the present work, quadrilateral plane stress elements (Q4 elements) with unit thickness and a fixed uniform mesh are used to discretize the design domain for structural analysis. Under this circumstance, the stiffness matrix of the $e$-th finite element can be calculated as

$$\boldsymbol{k}_e = \int_{\Omega_e} \boldsymbol{B}^{\top} \boldsymbol{D}_e \boldsymbol{B} \, \mathrm{dV} \tag{4.1}$$

where $\boldsymbol{B}$ is the strain-displacement matrix, $\Omega_e$ represents the region occupied by the $e$-th finite element, and $\boldsymbol{D}_e = \rho_e \boldsymbol{D}^{\mathrm{s}}$ with $\boldsymbol{D}^{\mathrm{s}}$ and $\rho_e$ denoting the constitutive matrix of the solid material and the artificial density associated with the $e$-th element. In the present work, the artificial density $\rho_e$ of the element is interpolated as

$$\rho_e = \frac{\sum_{i=1}^{4} \left( H_\epsilon\left( (\phi^{\mathrm{s}})_i^e \right) \right)^q}{4}. \tag{4.2}$$

In Eq (4.2), $(\phi^{\mathrm{s}})_i^e$ is the value of TDF of the whole structure at the $i$-th node of element $e$, $q > 1$ is a penalization factor (in the present work, $q = 2$ is used) and $H_\epsilon(x)$ is a regularized Heaviside function for numerical implementation, which takes the following form:

$$H_\epsilon(x) = \begin{cases} 1, & \text{if } x > \epsilon, \\ \dfrac{3(1-\alpha)}{4}\left(\dfrac{x}{\epsilon} - \dfrac{x^3}{3\epsilon^3}\right) + \dfrac{1+\alpha}{2}, & \text{if} -\epsilon \le x \le \epsilon, \\ \alpha, & \text{otherwise}, \end{cases} \tag{4.3}$$

where $\epsilon$ and $\alpha$ are two small positive numbers used for controlling the length of the transition zone between the solid and void parts and avoid the singularity of the global stiffness matrix. In the present work, we choose $\epsilon = (2{\sim}4) \times \min(\Delta x, \Delta y)$ and $\alpha = 10^{-3}$, respectively. Here $\Delta x$ and $\Delta y$ are the sizes of the FE mesh along two coordinate directions, respectively.

Based on the above ersatz material model, the sensitivity of the structural compliance and volumes with respect to an arbitrary design variable $d$ can be expressed as:

$$\frac{\partial C}{\partial d} = -\boldsymbol{u}^{\mathrm{T}} \frac{\partial \boldsymbol{K}}{\partial d} \boldsymbol{u} = -\sum_{e=1}^{NE} \left( \sum_{i=1}^{4} \frac{q}{4} (H_\epsilon(\phi^{\mathrm{s}})_i^e)^{q-1} \frac{\partial H_\epsilon((\phi^{\mathrm{s}})_i^e)}{\partial d} \boldsymbol{u}_e^{\mathrm{T}} \boldsymbol{k}^{\mathrm{s}} \boldsymbol{u}_e \right), \tag{4.4}$$

$$\frac{\partial V}{\partial d} = \frac{1}{4} \sum_{e=1}^{NE} \sum_{i=1}^{4} \frac{\partial H_\epsilon((\phi^{\mathrm{s}})_i^e)}{\partial d}, \tag{4.5}$$

$$\frac{\partial V_{\mathrm{in}}}{\partial d} = \frac{1}{4} \sum_{e=1}^{NE} \sum_{i=1}^{4} \frac{\partial H_\epsilon((\phi_0^{\mathrm{ext}})_i^e)}{\partial d}, \tag{4.6}$$

where $NE$ is the total number of elements in the design domain; $\boldsymbol{u}$ is the displacement field vector, $\boldsymbol{u}_e$ is the elemental displacement vector associated with the $e$-th element; $\boldsymbol{K}$ and $\boldsymbol{k}^{\mathrm{s}}$ are the global stiffness matrix and the elemental stiffness matrix of a solid element, respectively.

To obtain $\partial H_\epsilon(\phi^{\mathrm{s}})/\partial d$ and $\partial H_\epsilon(\phi_0^{\mathrm{ext}})/\partial d$, in numerical implementation, the max/min operators are approximated by the following well-known K-S function [46]:

$$\chi = \ln\left(\sum_{i=1}^{n} \exp(l\chi^i)\right)/l, \tag{4.7}$$

where $\chi \approx \max(\chi^1, \dots, \chi^n)$ if $l$ is a relatively large positive number (e.g., $l_+ = 50$), and $\chi \approx \min(\chi^1, \dots, \chi^n)$ when $l$ is a relatively large negative number (e.g., $l_- = -50$).

For the $j-$th design variable of $k-$th component in each cell of the infill structure, through the chain rule, we have

$$\frac{\partial H_{\epsilon}(\phi^{\mathrm{s}})}{\partial d_j^k}=\frac{\partial H_{\epsilon}(\phi^{\mathrm{s}})}{\partial \phi^{\mathrm{s}}}\frac{\partial \phi^{\mathrm{s}}}{\partial \phi_1}\frac{\partial \phi_1}{\partial \phi^{\mathrm{g-s}}}\frac{\partial \phi^{\mathrm{g-s}}}{\partial \phi^k}\left(\sum_{l=1}^{ncell}\left(\frac{\partial \phi^k}{\partial \phi^{k,l}}\frac{\partial \phi^{k,l}}{\partial d_j^{k,l}}\right)\right), \tag{4.8}$$

where

$$\phi^{\mathrm{s}}=\frac{\ln\big(\exp(l_-\phi_0)+\exp(l_-\phi_1)\big)}{l_-}, \tag{4.9$a$}$$

$$\phi_1=\frac{\ln\big(\exp(-l_+\phi_0^{\mathrm{ext}})+\exp(l_+\,\phi^{\mathrm{g-s}})\big)}{l_+}, \tag{4.9$b$}$$

$$\phi^{\mathrm{g-s}}=\frac{\ln\left(\sum_{\mathrm{k=1}}^{nc}\exp\left(l_+\,\phi^k(\widetilde{\boldsymbol{x}})\right)\right)}{l_+}, \tag{4.9$c$}$$

$$\phi^k(\widetilde{\boldsymbol{x}})=\frac{\ln\left(\sum_{l=1}^{ncell}\exp\left(l_+\,\phi^{k,l}(\widetilde{\boldsymbol{x}})\right)\right)}{l_+}. \tag{4.9$d$}$$

Therefore, we have

$$\frac{\partial H_{\epsilon}(\phi^{\mathrm{s}})}{\partial d_j^k}=\frac{\partial H_{\epsilon}(\phi^{\mathrm{s}})}{\partial \phi^{\mathrm{s}}}\frac{\exp((l_- - l_+)\phi_1)}{\exp(l_-\phi^{\mathrm{s}})}\sum_{l=1}^{ncell}\left(\exp\left(l_+\,\phi^{k,l}(\widetilde{\boldsymbol{x}})\right)\frac{\partial \phi^{k,l}}{\partial d_j^{k,l}}\right). \tag{4.10}$$

In Eq. (4.10), $\partial H_{\epsilon}/\partial \phi^{\mathrm{s}}$ can be easily obtained based on Eq. (4.3), and the details of $\partial \phi^{k,l}/\partial d_j^{k,l}$ can be found in the Appendix. It is worth noting that, as the TDF $\phi_0^{\mathrm{ext}}$ is independent of the infill structure, the sensitivity $\partial H_{\epsilon}(\phi_0^{\mathrm{ext}})/\partial d_j^k$ is zero for the MMC related design variables.

Similarly, for the $j$-th design variable of $k$-th void of the shell structure, we have

$$\frac{\partial H_{\epsilon}(\phi^{\mathrm{s}})}{\partial d_j^k}=\frac{\partial H_{\epsilon}(\phi^{\mathrm{s}})}{\partial \phi^{\mathrm{s}}}\left(\frac{\partial \phi^{\mathrm{s}}}{\partial \phi_0}\frac{\partial \phi_0}{\partial \phi^k}\frac{\partial \phi^k}{\partial d_j^k}+\frac{\partial \phi^{\mathrm{s}}}{\partial \phi_1}\frac{\partial \phi_1}{\partial \phi_0^{\mathrm{ext}}}\frac{\partial \phi_0^{\mathrm{ext}}}{\partial \phi^{k-\mathrm{ext}}}\frac{\partial \phi^{k-\mathrm{ext}}}{\partial d_j^k}\right), \tag{4.11a}$$

$$\frac{\partial H_{\epsilon}(\phi_0^{\mathrm{ext}})}{\partial d_j^k}=\frac{\partial H_{\epsilon}(\phi_0^{\mathrm{ext}})}{\partial \phi_0^{\mathrm{ext}}}\frac{\partial \phi_0^{\mathrm{ext}}}{\partial \phi^{k-\mathrm{ext}}}\frac{\partial \phi^{k-\mathrm{ext}}}{\partial d_j^k}, \tag{4.11b}$$

where

$$\phi_0=\frac{\ln\left(\sum_{k=1}^{nv}\left(\exp(l_-\phi^k)\right)\right)}{l_-}, \tag{4.12$a$}$$

$$\phi_0^{\text{ext}} = \frac{\ln\left(\sum_{k=1}^{nv}\left(\exp(l_- \phi^{k-\text{ext}})\right)\right)}{l_-}, \qquad (4.12b)$$

with $\phi^k$ and $\phi^{k-\text{ext}}$ denoting the TDFs of $k$-th void and $k$-th expanded void, respectively.

Furthermore, it is easy to obtain

$$\frac{\partial H_\epsilon(\phi^{\text{s}})}{\partial d_j^k} = \frac{\partial H_\epsilon(\phi^{\text{s}})}{\partial \phi^{\text{s}}}\left(\frac{\exp(l_-\phi^k)}{\exp(l_-\phi^s)}\frac{\partial \phi^k}{\partial d_j^k} - \frac{\exp((l_- - l_+)\phi_1)\exp(l_-\phi^{k-\text{ext}})}{\exp(l_-\phi^s)\exp\left((l_+ + l_-)\phi_0^{\text{ext}}\right)}\frac{\partial \phi^{k-\text{ext}}}{\partial d_j^k}\right), \quad (4.13)$$

and

$$\frac{\partial H_\epsilon(\phi_0^{\text{ext}})}{\partial d_j^k} = \frac{\partial H_\epsilon(\phi_0^{\text{ext}})}{\partial \phi_0^{\text{ext}}}\frac{\exp(l_-\phi^{k-\text{ext}})}{\exp(l_-\phi_0^{\text{ext}})}\frac{\partial \phi^{k-\text{ext}}}{\partial d_j^k}. \qquad (4.14)$$

Since $\phi^k$ and $\phi^{k-\text{ext}}$ are explicit functions (B-splines equation) of the design variable $d_j^k$, the calculation of $\partial\phi^k/\partial d_j^k$ and $\partial\phi^{k-\text{ext}}/\partial d_j^k$ is trivial and will not be discussed in detail here. The interested readers are referred to the appendix of [35] for more details.

In the present study, unless otherwise stated, the center of each component in the prototype unit cell is fixed. Under this circumstance, we have $\overline{\boldsymbol{D}}_k^{\text{C}} = (a_k, \theta_k, t_k^1, t_k^2, (\boldsymbol{\alpha}^k)^\top, (\boldsymbol{\beta}^k)^\top)^\top$ for $k = 1, \dots, nc$. In other words, except for the coefficients of the CPFs, each component in the prototype unit cell only has 4 design variables. Besides, in this work, the number of B-spline control points is chosen as 12 for each void. Therefore, for each void, there are only 14 (2 for the coordinates of the central point) design variables. As shown in the numerical examples provided in Section 5, in the proposed approach, even a small number of design variables can describe the geometries of fairly complicated shell-graded-infill structures. This also renders the possibility of combing the proposed approach with gradient-free algorithms (e.g., genetic algorithm) to deal with complex problems where analytical sensitivity information is difficult (or even impossible) to be obtained.

## 5. Numerical examples

In this section, three plane stress examples are investigated to illustrate the effectiveness of the proposed method. Without loss of generality, it is assumed that all involved quantities are dimensionless and the thickness of all design domains takes a unit value. The Young's modulus and the Poisson's ratio of the isotropic solid material are chosen as $E^s = 1$ and $\nu^s = 0.3$, respectively. Besides, the design variables are updated by the MMA algorithm [47], and the optimization process is terminated if the maximum value of the relative change of each design variable between two consecutive iterations is below a specified threshold (i.e., 5%). Finally, quadrilateral plane stress elements are adopted for finite element analysis in all examples.

### *5.1 A short beam example*

In this example, the design domain, boundary conditions, and external load are shown in Fig. 11. It is intended to find the optimal material distribution minimizing the structural compliance, with constraints on both the total volume of solid material and the volume of the infill region. The maximum available volume of the solid material is $\bar{V} = 0.3V_{\mathrm{D}}$ with $V_{\mathrm{D}}$ denoting the volume of the design domain.

As illustrated by Fig. 12, for describing the geometry of graded lattice background it is assumed that the design domain is covered by 40 (along the width direction) by 24 (along the length direction) lattice cells, and two components are distributed in the prototype unit cell. Besides, in order to describe the geometry of the shell structure, eight internal "voids" are introduced into the initial design. To further enhance the flexibility of the variation of structural configuration, the outer boundary of the initial design is also represented by a B-spline curve. The thickness of all the coating shell is assumed to be uniform and takes a fixed value as $\Delta d = 0.08$. Furthermore, in order to demonstrate the flexibility of the B-spline curve for geometry representation, the initial structure is deliberately separated from the clamped boundary and weak material ($E^w = 1 \times 10^{-6}$) is used to avoid the singularity of the stiffness matrix. During the optimization process, a uniform 500×300 mesh is used for finite element analysis.

The adopted global CPFs for generating the graded lattice structure are expanded using the first 4 terms of a trigonometric basis as described in Eq. (2.14) (i.e., $n_1 = n_2 = 4$). Under this circumstance, there are only 150 design variables (i.e., 4×2=8 for geometry parameters of the two components, 8×2=16 for the CPFs, and 14×9=126 for the B-spline boundaries, respectively) involved in the optimization problem.

Firstly, the optimization problem is tested without the volume constraint imposed on the infill region. As illustrated in Fig. 13, the optimized structure is composed of non-porous solid infill. The underlying reason is that, as pointed out by Yan et al. [48], for pure minimum compliance design under a volume constraint of available solid material, the optimal design is composed of non-porous solid microstructures. This result, to a certain extent, validates the proposed algorithm and also illustrates the necessity of introducing a lower bound constraint on the volume of the infill region in order to generate a lattice structure when structural compliance is minimized.

Fig. 14a-Fig. 14d show the optimized shell structures with graded infill lattice when $\underline{V}$ takes the value of $0.5V_{\mathrm{D}}$, $0.6V_{\mathrm{D}}$, $0.7V_{\mathrm{D}}$, $0.8V_{\mathrm{D}}$, respectively. It can be observed clearly that all optimized structures are composed of graded infills varying along different directions. Meanwhile, the connection between adjacent unit cells are natural and smooth with the help of the proposed coordinate perturbation technique.

In addition, it is also found that as the value of $\underline{V}$ increases (i.e., from $0.5V_{\mathrm{D}}$ to $0.8V_{\mathrm{D}}$), the volume of the internal voids enclosed by the shells is decreased, and the optimized value of structural compliance is also increased (i.e., from 90.06 to 122.25). This is consistent with the observations in Yan et al. [48] and Wu et al. [24] that the lattice-like structures are actually inferior to the structures without porous microstructures when only the overall stiffness is considered. Nevertheless, it is worth noting that the shell structures with lattice infills are in general more efficient to preserve structural stability under compressive load [8], more robust under load/material uncertainty [9] and have much better energy absorption capability [49]. Since the main purpose of the present work is to develop an explicit approach to optimize the shell-graded-infill structures, only the compliance minimization problem is studied. It can be expected the proposed approach can also find its application when the structural stability, load/material uncertainty, and other porous microstructure-friendly structural performance measures are considered.

On the other hand, the optimized structures contain complex structural details with small feature sizes, a natural question is whether the mesh used is fine enough to obtain meaningful optimized designs. To answer this question, the optimized structures are reanalyzed by adopting gradually refined FE meshes. As shown in Fig. 15, when the finite elements are as dense as 150 per unit length (which corresponds to a 3000×1800 mesh for the whole design domain), the compliance of the four optimized structures are 116.67, 161.08, 180.26, 203.70, respectively. Although the FE analysis results may not be the converged values (with respect to the corresponding theoretical solutions), it is found that the rank of the compliance values of the optimized designs keeps invariant during the refinement of meshes as Fig. 14. This fact is consistent with the conclusion obtained in the studies on approximate reanalysis [50], and more specifically, it seems reasonable to claim that the topology of the optimized structure is insensitive to the density of the adopted FE mesh.

In addition, it is also observed that the topologies of the optimized structures are similar to the optimized cantilever beams obtained in [24]. Moreover, even performing FE analysis by adopting a 3000×1800 mesh, the compliances of the optimized structures of the proposed method are still comparable with the results in [24]. Fig. 16 plots a representative iteration curve, which shows that the optimization process is relatively stable.

Fig. 17 presents the optimized shell-uniform-infill design with $\underline{V} = 0.5V_{\mathrm{D}}$ and $\Delta d = 0.08$. With the use of a relatively coarse 500×300 mesh, the optimized structural compliance decreases from 134.15 to 90.06 with a relative reduction of 32.9% (when a refined 3000×1800 mesh is adopted, the corresponding values are 176.50, 116.67 and 33.9%) by introducing the graded pattern into the infill structure, which illustrates clearly the advantage of shell-graded-infill designs.

### *5.2 A MBB example*

The problem setting is described schematically in Fig. 18. Specifically, a unit vertical load $F = 1$ is imposed on the middle point of the top side. The thickness of the coating shell is fixed as $\Delta d = 0.06$ and the upper bound of the volume of available solid material is $\bar{V} = 0.3V_{\mathrm{D}}$. Meanwhile, in the following four tested cases, the lower bounds of the volume of the infill region (i.e., $\underline{V}$) is chosen as $0.5V_{\mathrm{D}}$, $0.6V_{\mathrm{D}}$, $0.7V_{\mathrm{D}}$, $0.8V_{\mathrm{D}}$, respectively.

The same initial design is used for all considered cases. As shown in Fig. 19, the design domain contains 64 (along the width direction) by 16 (along the length direction) lattice cells, with two components in the prototype unit cell. The 12 internal voids and the outer boundary of the initial design are all represented by B-spline curves. In addition, a uniform 960×240 mesh is adopted for finite element analysis.

With the use of similar global CPFs as in the short beam example, the optimized structures for the four considered cases are shown in Fig. 20a-Fig. 20d, respectively. It can be seen from these optimization results that the optimized structural performances share the same trend as in the above example, i.e., the larger the volume of the infill region, the smaller the overall stiffness (measured in terms of the structural compliance) of the optimized structure. Fig. 21 presents the convergence curves of the structural compliance with respect to different refinements of FE mesh. Similarly, the rank of the compliances of the four optimized structures stays the same with respect to different FE meshes.

It is also worth noting that thanks to the explicit description of structural geometry in the MMC- and MMV-based approaches, the optimized designs can be easily transferred to CAD/CAE system to build the geometry model without any post-processing. Table 1 lists the coordinates of the control points associated with the outer boundary indicated by $c_1$ shown in Fig. 20a. Moreover, the optimized shell-graded-infill structures are obtained with the use of only a small number of design variables (actually 4×2+8×2+14×13=206). These two features, however, in general cannot be easily achieved by traditional implicit geometry description-based approaches.

### *5.3 A multi-load example*

The last example is a multi-load problem and the initial design dependency of the proposed method is also tested. Fig. 22 plots the design domain, boundary conditions, and external loads. The upper bound of the available solid material volume is $\bar{V} = 0.3V_D$ and the lower bound of the volume of the infill region is chosen as $\underline{V} = 0.5V_\mathrm{D}$. For all cases, a uniform 600×300 meshes is adopted for finite element analysis.

Firstly, the optimization problem is considered for the single load case (i.e., $F_1, F_2, F_3$ are applied simultaneously). Three initial designs with different numbers of voids are adopted, as shown in Fig. 23a-Fig.23c. For all the three considered cases, the global CPFs are chosen in the form described in Eq. (2.14) with parameters $n_1 = 4$ and $n_2 = 4$. Fig. 24a-Fig. 24c plot the corresponding optimized structures. It is found that for the 8 initial voids case, the optimized structural topology is relatively simple, and the structural compliance is 243.67, which is larger than that of the other two cases. While for the 12 and 18 initial voids cases, the optimized topologies are similar and the corresponding compliances are also very close (i.e., 232.68 and 232.03). This indicates that the number of initial holes does have some effect on the final optimized designs since the considered problem is highly non-convex, however, this effect can be alleviated to some extent by introducing more holes in the initial design.

The effect of the number of basis functions used in CPFs is also tested for the single load case. Fig. 25 shows the optimized design with respect to the initial design shown in Fig. 23c and the CPF parameters are chosen as $n_1 = n_2 = 2$. The corresponding structural compliance is 243.08, which is 4.8% larger than that of the optimized design obtained with $n_1 = n_2 = 4$ (shown in Fig. 24c). This result is reasonable since more numbers of CPF basis may lead to a larger design space.

Besides, in the above examples, the numbers of components in the prototype unit cell are two and the central points of the components are fixed during the optimization process. It is necessary to test the effect of a more complex prototype unit cell on the optimization results. Fig. 26 shows a different initial design with four components in the prototype unit cell, and the central points of these components are allowed to move within a small range $x_{1,1}^k \in [0.49, 0.51]$, $y_{1,1}^k \in [0.49, 0.51]$, with $k = 1, 2, 3, 4$. To avoid the optimized structure have too many small details, the sizes of the design domain are reduced by a factor of two, and the coating thickness is reduced to $\Delta d = 0.05$. The total number of components in the design domain is the same as in Fig. 23c. Moreover, the CPF parameters are chosen as $n_1 = 4$ and $n_2 = 4$. The corresponding optimized structure is shown in Fig. 27. It is found that a more complex (rectangular-shaped and triangular-shaped) infill pattern appears in this optimized design and about 8.1% improvement of the compliance is achieved as compared with the design shown in Fig. 24c.

In addition, it is believed that the triangular-shaped microstructures are optimal subjected to multi-load cases [24]. Therefore, it can be expected that the optimized shell-graded-infill structures could gain more benefits in multi-load problems. Fig. 28a and Fig. 28b present the optimized structures of the multi-load cases, i.e., the three vertical loads applied individually, and the objective function is the average of three compliances corresponding to each individual load. The initial designs of these two cases are the same as the example in Fig. 23c and Fig. 26, respectively. It is interesting to find that the optimized structure illustrated Fig. 28b has triangular-shaped infills and achieves an almost 14.3% improvement of the compliance compared with the optimized structure in Fig. 28a. This is quite consistent with our expectation. It is also realized that for more complex infill structures, a finer finite element mesh and an increased amount of computation cost are desired. Therefore, it is also necessary to introduce the graded-microstructure oriented homogenization technique to reduce the computational cost [51], which is one of our future research directions.

## 6. Concluding remarks

In the present work, a new approach for designing coated structures with graded infill is developed with the use of a hybrid MMC-MMV approach. The key idea is to describe the geometry of the shell-graded-infill structures using a set of geometry parameters. Under such treatment, both the crisp boundary of the coating shell and the graded infill can be optimized simultaneously, without any post-processing, only using a very small number of design variables. To control the material distribution in the infill region, a constraint on the volume of the infill region is further introduced. Compared with the optimized designs of coated structures with uniform infill, it is found that the structural stiffness can be greatly improved through the proposed concurrent design approach for shell-graded-infill structures.

It is also worth noting that, as shown in [5, 30], the explicit description of the structural geometry is very helpful for dealing with manufacturing-oriented constraints (e.g., controlling minimum feature size or overhang angles of structural components) which should also be considered during the design of shell-graded-infill structures for additive manufacturing. Moreover, although in the present work, the corresponding design and optimization is only carried out under the deterministic framework, the developed methodology can also be applied to take the influence of possible uncertainties (e.g., induced by the manufacturing process) into consideration. For instance, the geometrical uncertainties can be conveniently represented by introducing a dimensional tolerance on structural components. Based on the present work, future research directions deserve intensive explorations including the optimal design of three-dimensional coated structures with graded lattice infill considering manufacturing constraints/uncertainties, buckling constraint and multi-physics requirements, etc. Corresponding results will be reported in separated works.

## Appendix: The sensitivities of the coordinate-transformed TDF of the component with respect to the graded-infill related design variables

The aim of this section is to calculate the $\partial\phi^{k,l}/\partial d_j^{k,l}$ in Eq. (4.10). For the $k$-th component in each unit cell, the design variable vector is $\overline{\boldsymbol{D}}_k^{\mathrm{C}} = \left(x^{k,1}, y^{k,1}, a_k, \theta_k, t_k^1, t_k^2, \left(\boldsymbol{\alpha}^k\right)^\top, \left(\boldsymbol{\beta}^k\right)^\top\right)^\top$, with $\boldsymbol{\alpha}^k = \left(\alpha_{1,1}^k, \alpha_{1,2}^k, \dots, \alpha_{r,1}^k, \alpha_{r,2}^k, \dots, \alpha_{n_1,1}^k, \alpha_{n_1,2}^k\right)^\top$, $\boldsymbol{\beta}^k = \left(\beta_{1,1}^k,\ \beta_{1,2}^k, \dots, \beta_{t,1}^k,\ \beta_{t,2}^k, \dots, \beta_{n_2,1}^k, \beta_{n_2,2}^k\right)^\top$ denoting the vectors of the coefficients associated with the perturbation basis functions along two coordinate directions.

According to the chain rule, we have：

$$\begin{cases}
\dfrac{d\phi^{k,l}}{dx^{k,1}} = \dfrac{\partial\phi^{k,l}}{\partial x'}\dfrac{\partial x'}{\partial x^{k,1}} + \dfrac{\partial\phi^{k,l}}{\partial y'}\dfrac{\partial y'}{\partial x^{k,1}} + \dfrac{\partial\phi^{k,l}}{\partial b_k}\dfrac{\partial b_k}{\partial x'}\dfrac{\partial x'}{\partial x^{k,1}}, \\
\dfrac{d\phi^{k,l}}{dy^{k,1}} = \dfrac{\partial\phi^{k,l}}{\partial x'}\dfrac{\partial x'}{\partial y^{k,1}} + \dfrac{\partial\phi^{k,l}}{\partial y'}\dfrac{\partial y'}{\partial y^{k,1}} + \dfrac{\partial\phi^{k,l}}{\partial b_k}\dfrac{\partial b_k}{\partial x'}\dfrac{\partial x'}{\partial y^{k,1}}, \\
\dfrac{d\phi^{k,l}}{da_k} = \dfrac{\partial\phi^{k,l}}{\partial a_k} + \dfrac{\partial\phi^{k,l}}{\partial b_k}\dfrac{\partial b_k}{\partial a_k}, \\
\dfrac{d\phi^{k,l}}{dt_k^1} = \dfrac{\partial\phi^{k,l}}{\partial b_k}\dfrac{\partial b_k}{\partial t_k^1}, \\
\dfrac{d\phi^{k,l}}{dt_k^2} = \dfrac{\partial\phi^{k,l}}{\partial b_k}\dfrac{\partial b_k}{\partial t_k^2}, \\
\dfrac{d\phi^{k,l}}{d\theta_k} = \dfrac{\partial\phi^{k,l}}{\partial x'}\dfrac{\partial x'}{\partial \theta_k} + \dfrac{\partial\phi^{k,l}}{\partial y'}\dfrac{\partial y'}{\partial \theta_k} + \dfrac{\partial\phi^{k,l}}{\partial b_k}\dfrac{\partial b_k}{\partial x'}\dfrac{\partial x'}{\partial \theta_k}, \\
\dfrac{d\phi^{k,l}}{d\alpha_{r,i}^k} = \dfrac{\partial\phi^{k,l}}{\partial x'}\dfrac{\partial x'}{\partial \tilde{x}}\dfrac{\partial \tilde{x}}{\partial \alpha_{r,i}^k} + \dfrac{\partial\phi^{k,l}}{\partial y'}\dfrac{\partial y'}{\partial \tilde{x}}\dfrac{\partial \tilde{x}}{\partial \alpha_{r,i}^k} + \dfrac{\partial\phi^{k,l}}{\partial b_k}\dfrac{\partial b_k}{\partial x'}\dfrac{\partial x'}{\partial \tilde{x}}\dfrac{\partial \tilde{x}}{\partial \alpha_{r,i}^k}, r = 1, \dots, n_1, i = 1, 2, \\
\dfrac{d\phi^{k,l}}{d\beta_{t,j}^k} = \dfrac{\partial\phi^{k,l}}{\partial x'}\dfrac{\partial x'}{\partial \tilde{y}}\dfrac{\partial \tilde{y}}{\partial \beta_{t,j}^k} + \dfrac{\partial\phi^{k,l}}{\partial y'}\dfrac{\partial y'}{\partial \tilde{y}}\dfrac{\partial \tilde{y}}{\partial \beta_{t,j}^k} + \dfrac{\partial\phi^{k,l}}{\partial b_k}\dfrac{\partial b_k}{\partial x'}\dfrac{\partial x'}{\partial \tilde{y}}\dfrac{\partial \tilde{y}}{\partial \beta_{t,j}^k}, t = 1, \dots, n_2, j = 1,2,
\end{cases}$$

where

$$\begin{cases} \dfrac{\partial b_k}{\partial x'} = \dfrac{t_k^2 - t_k^1}{2a_k}, \\ \dfrac{\partial b_k}{\partial a_k} = -\dfrac{t_k^2 - t_k^1}{2} x' a_k^{-2}, \\ \dfrac{\partial b_k}{\partial t_k^1} = \dfrac{1}{2} - \dfrac{x'}{2a_k}, \\ \dfrac{\partial b_k}{\partial t_k^2} = \dfrac{1}{2} + \dfrac{x'}{2a_k}, \end{cases}$$

$$\begin{cases} \dfrac{\partial \phi^{k,l}}{\partial x'} = -\left[\left(\dfrac{x'}{a_k}\right)^p + \left(\dfrac{y'}{b_k}\right)^p\right]^{\frac{1}{p}-1} x'^{p-1} a_k^{-p}, \\ \dfrac{\partial \phi^{k,l}}{\partial y'} = -\left[\left(\dfrac{x'}{a_k}\right)^p + \left(\dfrac{y'}{b_k}\right)^p\right]^{\frac{1}{p}-1} y'^{p-1} b_k^{-p}, \\ \dfrac{\partial \phi^{k,l}}{\partial a_k} = \left[\left(\dfrac{x'}{a_k}\right)^p + \left(\dfrac{y'}{b_k}\right)^p\right]^{\frac{1}{p}-1} x'^p {a_k}^{-p-1}, \\ \dfrac{\partial \phi^{k,l}}{\partial b_k} = \left[\left(\dfrac{x'}{a_k}\right)^p + \left(\dfrac{y'}{b_k}\right)^p\right]^{\frac{1}{p}-1} y'^p {b_k}^{-p-1}, \end{cases}$$

and

$$\begin{cases} \dfrac{\partial x'}{\partial x^{k,1}} = -\cos\theta_k, \\ \dfrac{\partial y'}{\partial x^{k,1}} = \sin\theta_k, \\ \dfrac{\partial x'}{\partial y^{k,1}} = -\sin\theta_k, \\ \dfrac{\partial y'}{\partial y^{k,1}} = -\cos\theta_k, \\ \dfrac{\partial x'}{\partial \theta_k} = -\sin\theta_k\,(\tilde{x} - x^{k,l}) + \cos\theta_k\,(\tilde{y} - y^{k,l}), \\ \dfrac{\partial y'}{\partial \theta_k} = -\cos\theta_k\,(\tilde{x} - x^{k,l}) - \sin\theta_k\,(\tilde{y} - y^{k,l}), \end{cases}$$

$$
\begin{cases}
\dfrac{\partial x'}{\partial \tilde{x}} = \cos\theta_k, \\
\dfrac{\partial y'}{\partial \tilde{x}} = -\sin\theta_k, \\
\dfrac{\partial x'}{\partial \tilde{y}} = \sin\theta_k, \\
\dfrac{\partial y'}{\partial \tilde{y}} = \cos\theta_k, \\
\dfrac{\partial \tilde{x}}{\partial \alpha_{r,1}^k} = \cos\left(\dfrac{(r-1)\pi}{L}\left(x - \dfrac{L}{2}\right)\right), r = 1, \dots, n_1, \\
\dfrac{\partial \tilde{x}}{\partial \alpha_{r,2}^k} = \sin\left(\dfrac{(r-1)\pi}{L}\left(x - \dfrac{L}{2}\right)\right), r = 1, \dots, n_1, \\
\dfrac{\partial \tilde{y}}{\partial \beta_{t,1}^k} = \cos\left(\dfrac{(t-1)\pi}{H}\left(y - \dfrac{H}{2}\right)\right), t = 1, \dots, n_2, \\
\dfrac{\partial \tilde{y}}{\partial \beta_{t,2}^k} = \sin\left(\dfrac{(t-1)\pi}{H}\left(y - \dfrac{H}{2}\right)\right), t = 1, \dots, n_2.
\end{cases}
$$

---

## Acknowledgements

The financial supports from the National Key Research and Development Plan (2016YFB0201600, 2016YFB0201601), the National Natural Science Foundation (11732004, 11821202), Program for Changjiang Scholars, Innovative Research Team in University (PCSIRT) and 111 Project (B14013) are also gratefully acknowledged.

## References

[1] M. Leary, L. Merli, F. Torti, M. Mazur, M. Brandt, Optimal topology for additive manufacture: A method for enabling additive manufacture of support-free optimal structures, Mater. Des. 63 (2014) 678–690.

[2] A.T. Gaynor, J.K. Guest, Topology optimization considering overhang constraints: Eliminating sacrificial support material in additive manufacturing through design, Struct. Multidiscip. Optim. 54 (2016) 1157–1172.

[3] M. Langelaar, An additive manufacturing filter for topology optimization of print-ready designs, Struct. Multidiscip. Optim. 55 (2017) 871–883.

[4] X. Qian, Undercut and overhang angle control in topology optimization: A density gradient based integral approach, Int. J. Numer. Methods Eng. 111 (2017) 247–272.

[5] X. Guo, J.H. Zhou, W.S. Zhang, Z.L. Du, C. Liu, Y. Liu, Self-supporting structure design in additive manufacturing through explicit topology optimization, Comput. Methods Appl. Mech. Engrg. 323 (2017) 27–63.

[6] G. Allaire, B. Bogosel, Optimizing supports for additive manufacturing, Struct. Multidiscip. Optim. 58 (2018) 2493–2515.

[7] J. Liu, A.T. Gaynor, S. Chen, Z. Kang, K. Suresh, A. Takezawa, L. Li, J. Kato, J. Tang, C.C.L. Wang, L. Cheng, X. Liang, Albert.C. To, Current and future trends in topology optimization for additive manufacturing, Struct. Multidiscip. Optim. 57 (2018) 2457–2483.

[8] A. Clausen, N. Aage, O. Sigmund, Exploiting additive manufacturing infill in topology optimization for improved buckling load, Engineering 2 (2016) 250–257.

[9] J. Wu, N. Aage, R. Westermann, O. Sigmund, Infill optimization for additive manufacturing—approaching bone-like porous structures, IEEE Trans. Vis. Comput. Graph. 24 (2018) 1127–1140.

[10] L. Osorio, E. Trujillo, A.W. Van Vuure, I. Verpoest, Morphological aspects and mechanical properties of single bamboo fibers and flexural characterization of bamboo/epoxy composites, J. Reinf. Plast. Compos. 30 (2011) 396–408.

[11] Y.J. Luo, Q. Li, S.T. Liu, Topology optimization of shell–infill structures using an

erosion-based interface identification method, Comput. Methods Appl. Mech. Engrg. 355 (2019) 94–112.

[12] Z. Liu, M.A. Meyers, Z. Zhang, R.O. Ritchie, Functional gradients and heterogeneities in biological materials: Design principles, functions, and bioinspired applications, Prog. Mater. Sci. 88 (2017) 467–498.

[13] S.W. Zhou, Q. Li, Design of graded two-phase microstructures for tailored elasticity gradients, J. Mater. Sci. 43 (2008) 5157-5167.

[14] D. Lin, Q. Li, W. Li, S.W Zhou, M.V. Swain, Design optimization of functionally graded dental implant for bone remodeling, Compos. Part. B-Eng. 40 (2009) 668-675.

[15] A. Radman, X. Huang, Y.M. Xie, Topology optimization of functionally graded cellular materials, J. Mater. Sci. 48 (2012) 1503-1510.

[16] A. Radman, X. Huang, Y.M. Xie, Maximizing stiffness of functionally graded materials with prescribed variation of thermal conductivity, Comput. Mater. Sci. 82 (2014) 457-463.

[17] J. Wu, C. Dick, R. Westermann, A system for high-resolution topology optimization, IEEE Trans. Vis. Comput. Graph. 22 (2016) 1195-1208.

[18] Y.Q. Wang, F.F. Chen, M.Y. Wang, Concurrent design with connectable graded microstructures, Comput. Methods Appl. Mech. Engrg. 317 (2017) 84-101.

[19] E. Garner, H.M.A. Kolken, C.C.L. Wang, A.A. Zadpoor, J. Wu, Compatibility in microstructural optimization for additive manufacturing, Addit. Manuf. 26 (2019) 65–75.

[20] A. Clausen, N. Aage, O. Sigmund, Topology optimization of coated structures and material interface problems, Comput. Methods Appl. Mech. Engrg. 290 (2015) 524–541.

[21] E. Wadbro, B. Niu, Multiscale design for additive manufactured structures with solid coating and periodic infill pattern, Comput. Methods Appl. Mech. Engrg. 357 (2019) 112605.

[22] Y.G. Wang, Z. Kang, A level set method for shape and topology optimization of coated structures, Comput. Methods Appl. Mech. Engrg. 329 (2018) 553–574.

[23] J. Fu, H. Li, M. Xiao, L. Gao, S. Chu, Topology optimization of shell-infill structures using a distance regularized parametric level-set method, Struct. Multidiscip. Optim. 59 (2019) 249–262.
[24] J. Wu, A. Clausen, O. Sigmund, Minimum compliance topology optimization of shell–infill composites for additive manufacturing, Comput. Methods Appl. Mech. Engrg. 326 (2017) 358-375.
[25] J.P. Groen, J. Wu, O. Sigmund, Homogenization-based stiffness optimization and projection of 2D coated structures with orthotropic infill, Comput. Methods Appl. Mech. Engrg. 349 (2019) 722–742.
[26] Y.Q. Wang, L. Zhang, S. Daynes, H.Y. Zhang, S. Feih, M.Y. Wang, Design of graded lattice structure with optimized mesostructures for additive manufacturing, Mater. Des. 142 (2018) 114-123.
[27] C. Liu, Z.L. Du, W.S. Zhang, Y.C. Zhu, X. Guo, Additive manufacturing-oriented design of graded lattice structures through explicit topology optimization, Trans. ASME J. Appl. Mech. 84 (2017) 081008.
[28] X. Guo, W.S. Zhang, W.L. Zhong, Doing topology optimization explicitly and geometrically—a new moving morphable components based framework, Trans. ASME J. Appl. Mech. 81 (2014) 081009.
[29] W.S. Zhang, W.Y. Yang, J.H. Zhou, D. Li, X. Guo, Structural topology optimization through explicit boundary evolution, Trans. ASME J. Appl. Mech. 84 (2017) 011011.
[30] W.S. Zhang, D. Li, J. Zhang, X. Guo, Minimum length scale control in structural topology optimization based on the Moving Morphable Components (MMC) approach, Comput. Methods Appl. Mech. Engrg. 311 (2016) 327–355.
[31] V.-N. Hoang, G.-W. Jang, Topology optimization using moving morphable bars for versatile thickness control, Comput. Methods Appl. Mech. Engrg. 317 (2017) 153–173.
[32] X. Guo, W.S. Zhang, J. Zhang, J. Yuan, Explicit structural topology optimization based on moving morphable components (MMC) with curved skeletons, Comput. Methods Appl. Mech. Engrg. 310 (2016) 711-748.

[33] W.S. Zhang, J. Yuan, J. Zhang, X. Guo, A new topology optimization approach based on moving morphable momponents (MMC) and the ersatz material model, Struct. Multidiscip. Optim. 53 (2016) 1243-1260.
[34] W.S. Zhang, J. Zhang, X. Guo, Lagrangian description based topology optimization-a revival of shape optimization, Trans. ASME J. Appl. Mech. 83 (2016) 041010.
[35] W.S. Zhang, J.S. Chen, X.F. Zhu, J.H. Zhou, D.C. Xue, X. Lei, X. Guo, Explicit three dimensional topology optimization via Moving Morphable Void (MMV) approach, Comput. Methods Appl. Mech. Engrg. 322 (2017) 590–614.
[36] C. Liu, Y.C. Zhu, Z. Sun, D.D. Li, Z.L. Du, W.S. Zhang, X. Guo, An efficient moving morphable component (MMC)-based approach for multi-resolution topology optim ization, Struct. Multidiscip. Optim. 58 (2018) 2455-2479.
[37] R.Y. Xue, C. Liu, W.S. Zhang, Y.C. Zhu, S. Tang, Z.L. Du, X. Guo, Explicit structural topology optimization under finite deformation via Moving Morphable Void (MMV) approach, Comput. Methods Appl. Mech. Engrg. 344 (2019) 798-818.
[38] X. Lei, C. Liu, Z.L. Du, W.S. Zhang, X. Guo, Machine learning-driven real-time topology optimization under moving morphable component-based framework, Trans. ASME J. Appl. Mech. 86 (2019) 011004.
[39] J.A. Norato, B.K. Bell, D.A. Tortorelli, A geometry projection method for continuum-based topology optimization with discrete elements, Comput. Methods Appl. Mech. Engrg. 293 (2015) 306–327.
[40] W.B. Hou, Y.D. Gai, X.F. Zhu, X. Wang, C. Zhao, L.K. Xu, K. Jiang, P. Hu, Explicit isogeometric topology optimization using moving morphable components, Comput. Methods Appl. Mech. Engrg. 326 (2017) 694–712.
[41] M. Takalloozadeh, G.H. Yoon, Implementation of topological derivative in the moving morphable components approach, Finite Elem. Anal. Des.. 134 (2017) 16–26.
[42] M.P. Bendsøe, Optimal shape design as a material distribution problem, Struct. Optim. 1 (1989) 193-202.
[43] M. Zhou, G.I.N. Rozvany, The COC algorithm, part II: topological, geometrical

and generalized shape optimization, Comput. Methods Appl. Mech. Engrg. 89 (1991) 309-336.

[44] G. Allaire, F. Jouve, A.M. Toader, Structural optimization using sensitivity analysis and a level-set method, J. Comput. Phys. 194 (2004) 363-393.

[45] M.Y. Wang, X.M. Wang, D.M. Guo, A level set method for structural topology optimization, Comput. Methods Appl. Mech. Engrg. 192 (2003) 227-246.

[46] G. Kreisselmeier, R. Steinhauser, Systematic control design by optimizing a vector performance index, IFAC Proceedings Volumes 12 (1979) 113-117.

[47] K. Svanberg, The method of moving asymptotes—a new method for structural optimization, Int. J. Numer. Methods Eng. 24 (1987) 359-373.

[48] J. Yan, G.D. Cheng, L. Liu, A uniform optimum material based model for concurrent optimization of thermoelastic structures and materials, Int. J. Simul. Multidisci. Des. Optim. 2 (2009) 259-266.

[49] M.F.A. Lorna, J. Gibson, Cellular Solids: Structure and properties, Cambridge University Press 1999.

[50] O. Amir, O. Sigmund, On reducing computational effort in topology optimization: how far can we go?, Struct. Multidiscip. Optim. 44 (2011) 25–29.

[51] Y.C. Zhu, S.S. Li, Z.L. Du, C. Liu, X. Guo, W.S. Zhang, A novel asymptotic-analysis-based homogenisation approach towards fast design of infill graded microstructures, J. Mech. Phys. Solids 124 (2019) 612–633.

## Tables

| Control point index | Coordinates | Control point index | Coordinates |
|---|---|---|---|
| 1 | (5.3974, 2.3705) | 7 | (1.7537, 2.3705) |
| 2 | (4.7928, 2.6627) | 8 | (0.3382, 0.0909) |
| 3 | (4.5192, 2.7733) | 9 | (3.0171, 0.1715) |
| 4 | (4.2867, 2.7961) | 10 | (4.2867, 0.0682) |
| 5 | (3.7632, 3.2772) | 11 | (5.5678, 0.1516) |
| 6 | (2.0472, 3.6635) | 12 | (8.3374, 0.0318) |

Table 1. Coordinates (relative to the coordinate system shown in Fig. 18) of the control points for the B-spline boundary curve shown in Fig. 20a.

## Figures

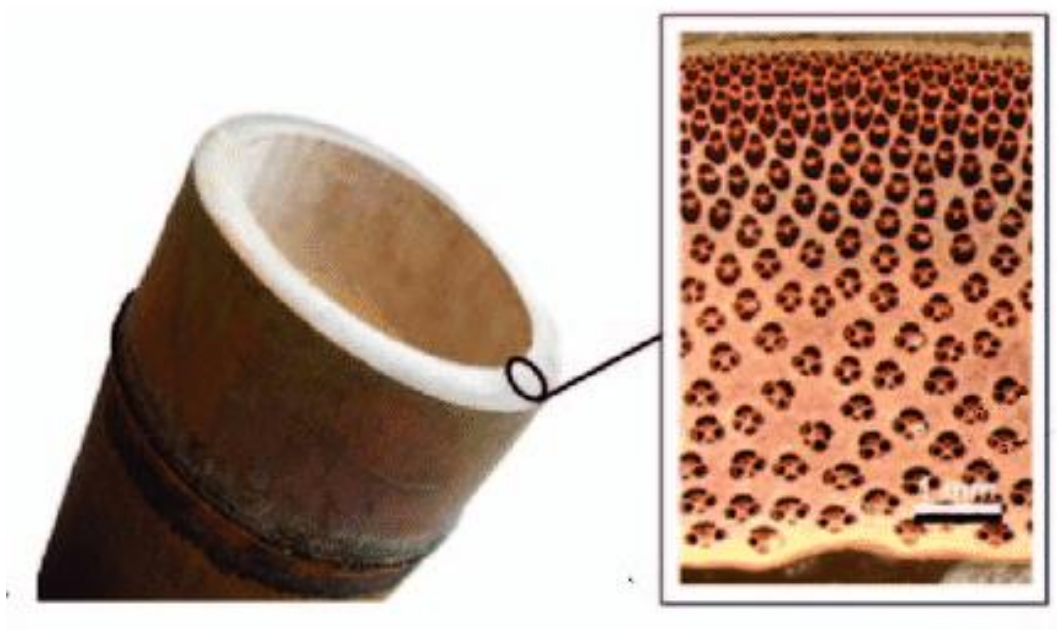

(a) Cross section of a bamboo [10].

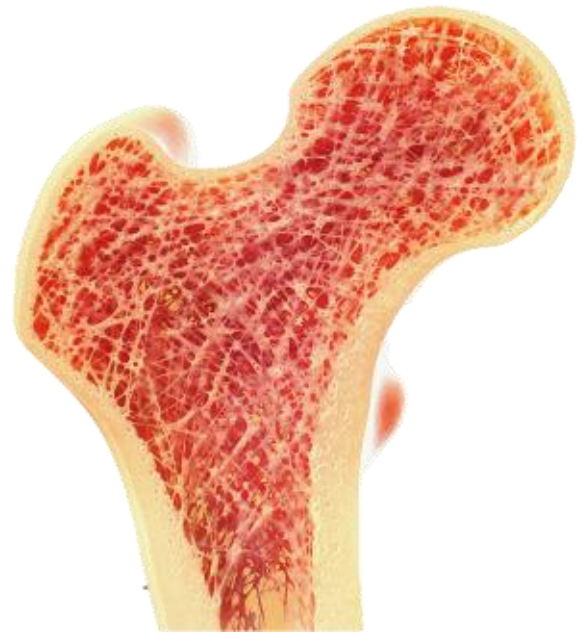

(b) A femur bone of human [11].

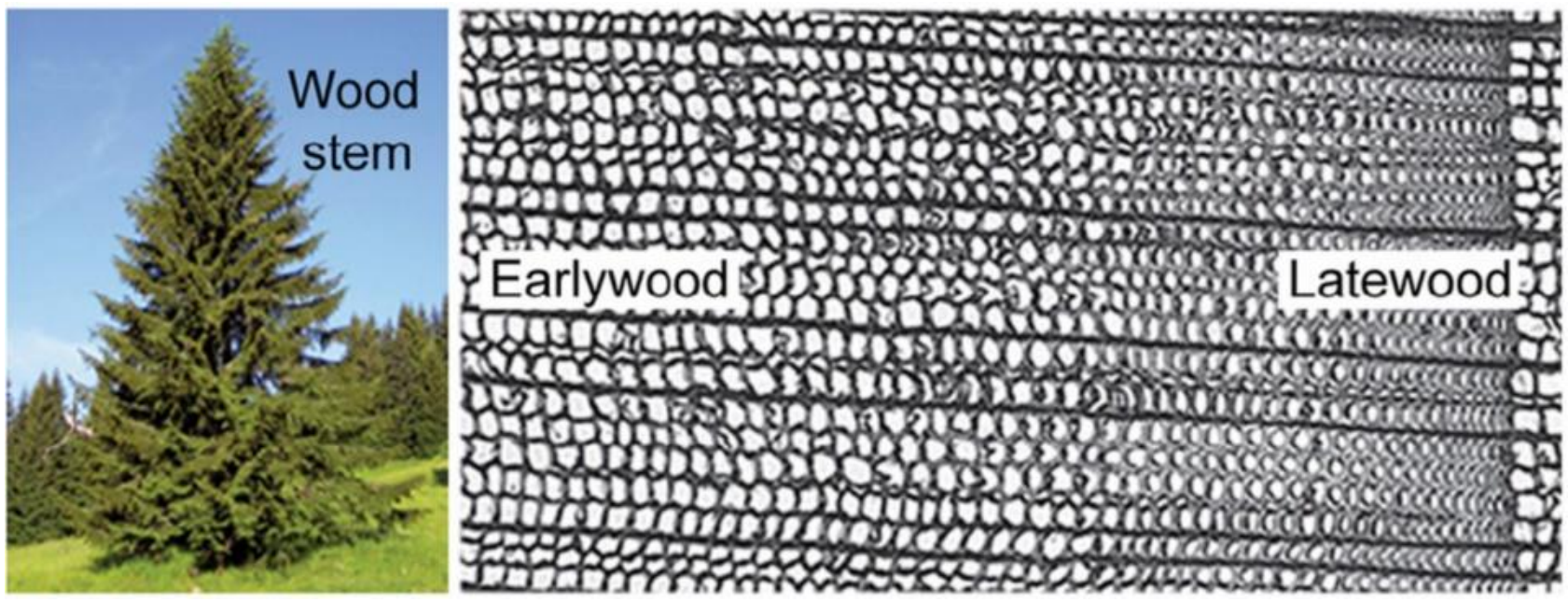

(c) The stem of the Norway spruce [12].

Fig. 1 Examples of shell–infill structures in nature.

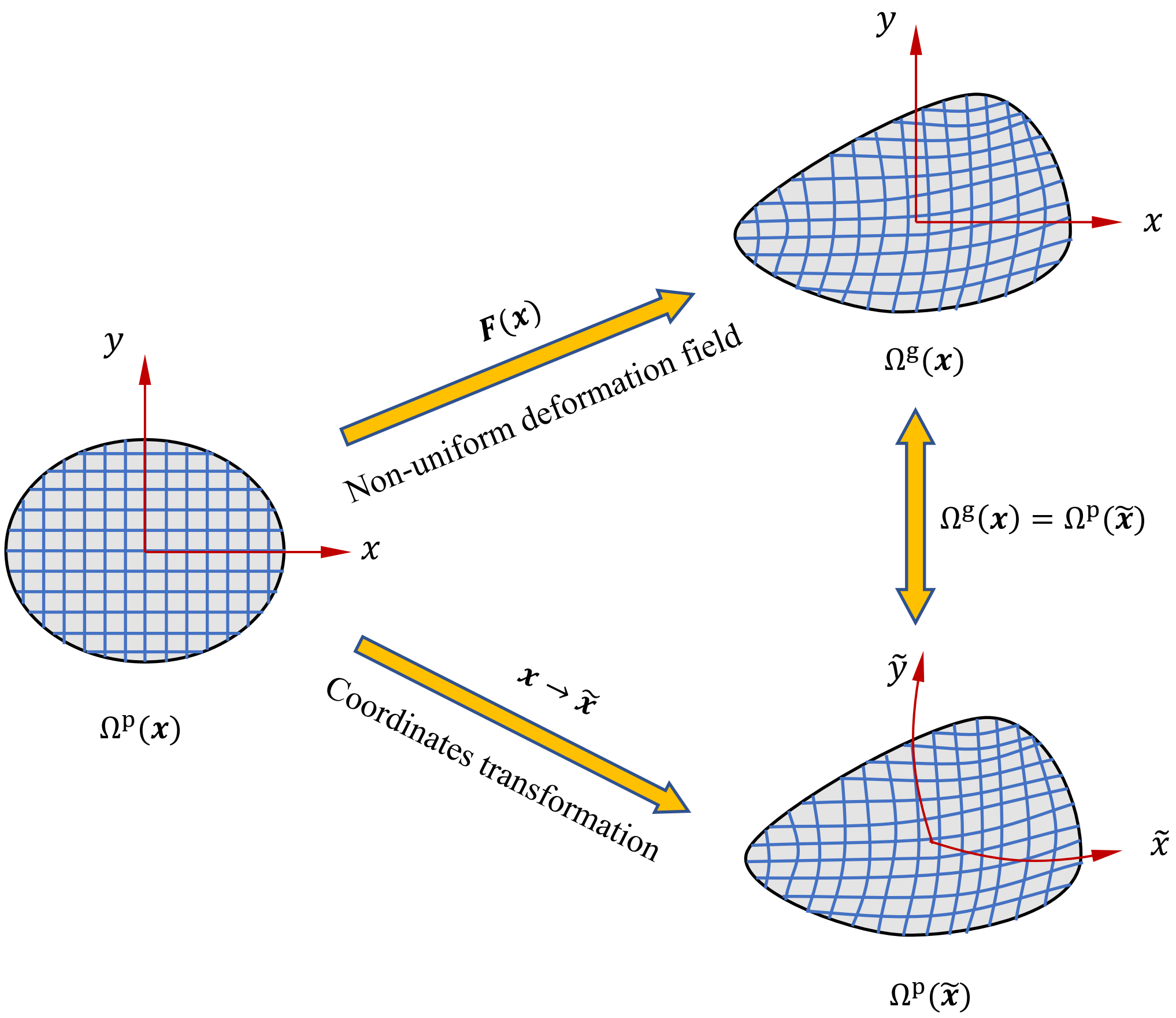

Fig. 2 Schematic diagram for the construction of a graded lattice by the perturbation of a uniform lattice structure.

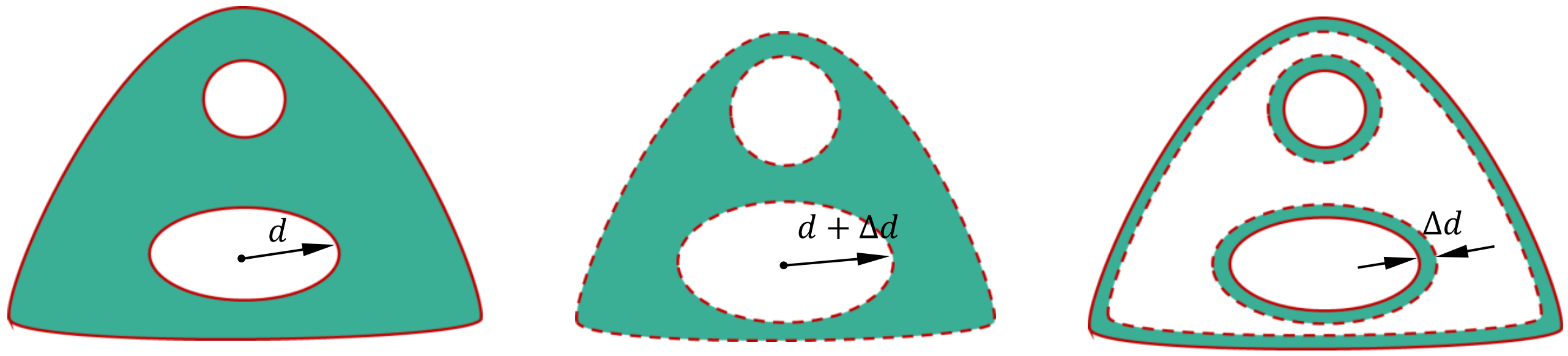

(a) $\Omega$ (b) $\Omega^{\text{ext}}$ (b) $\Omega^{\text{shell}} = \Omega \backslash \Omega^{\text{ext}}$

Fig. 3 A schematic diagram for the construction of a coating shell.

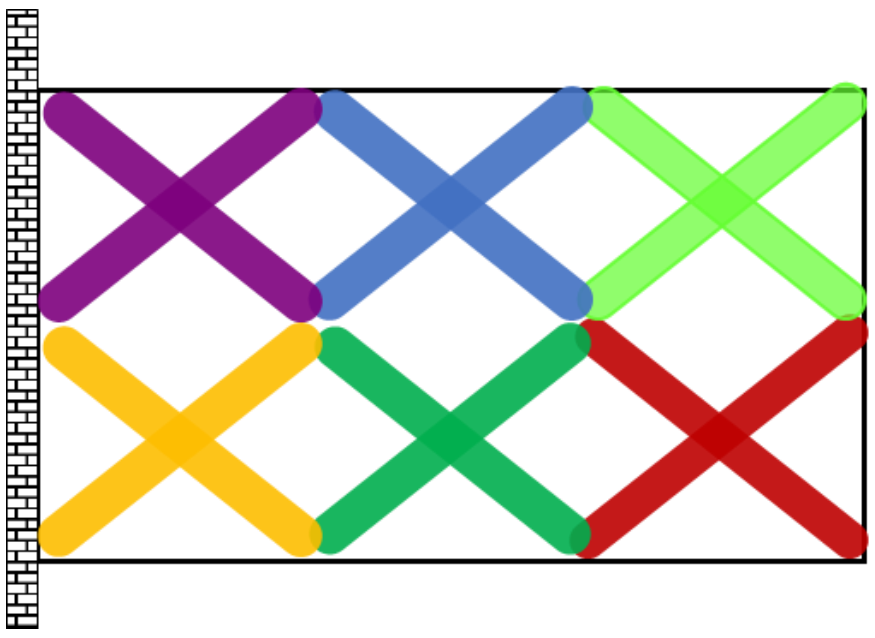

(a) The initial layout of the components.

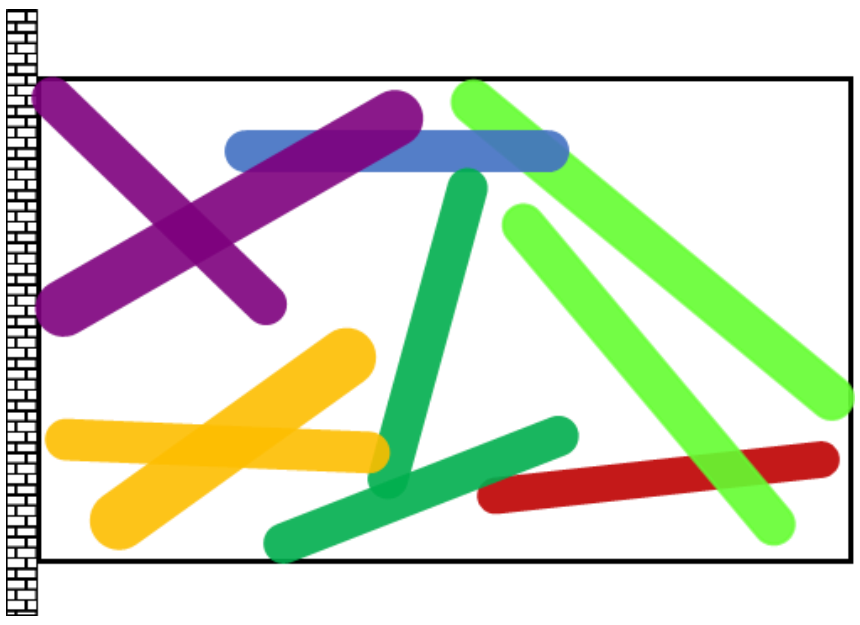

(b) Optimization process.

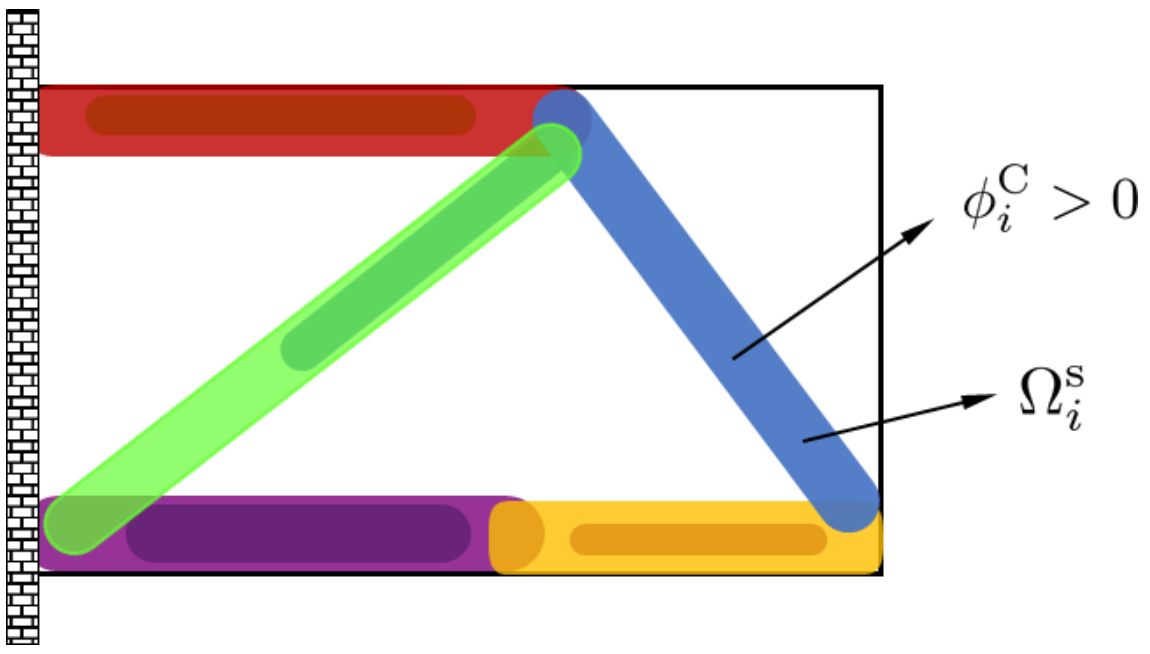

(c) The optimized layout of the components.

Fig. 4 A schematic illustration of the MMC-based topology optimization method.

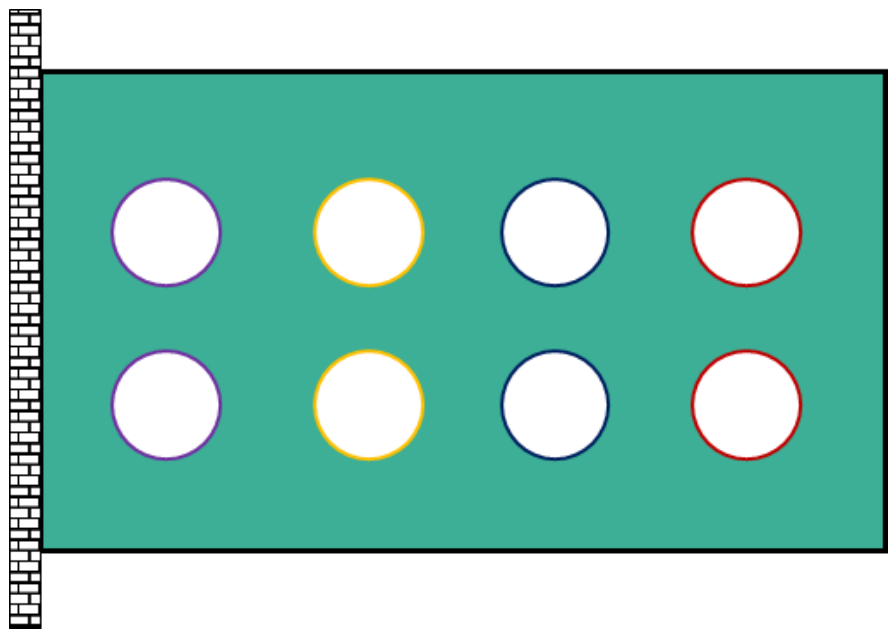

(a) The initial layout of the voids.

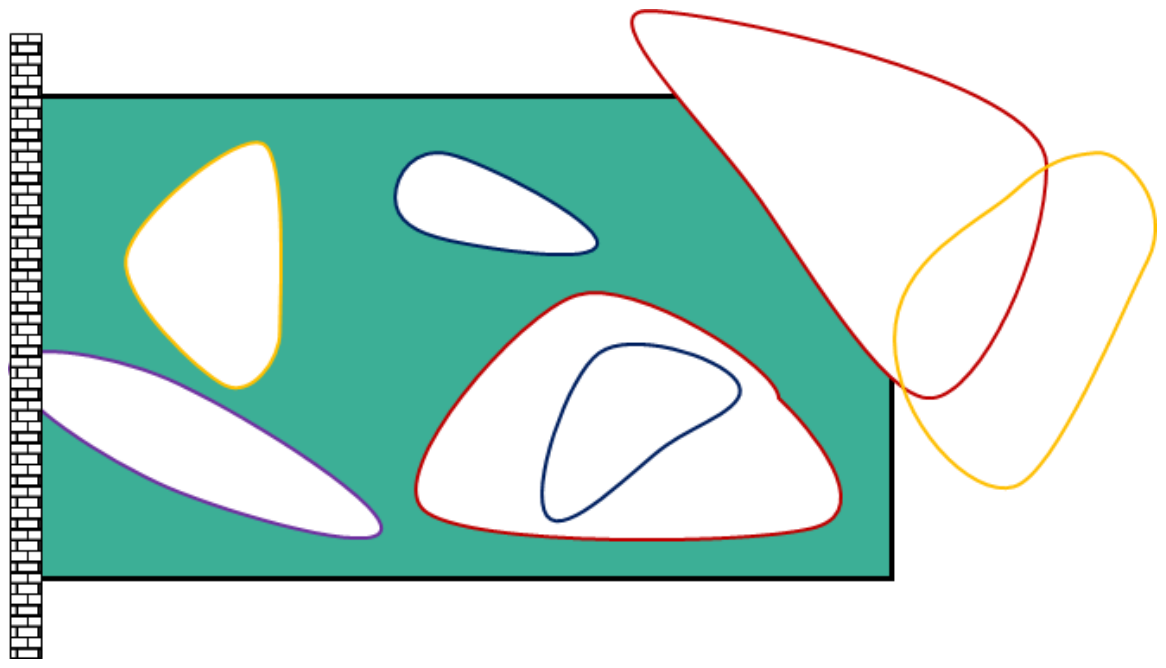

(b) Optimization process.

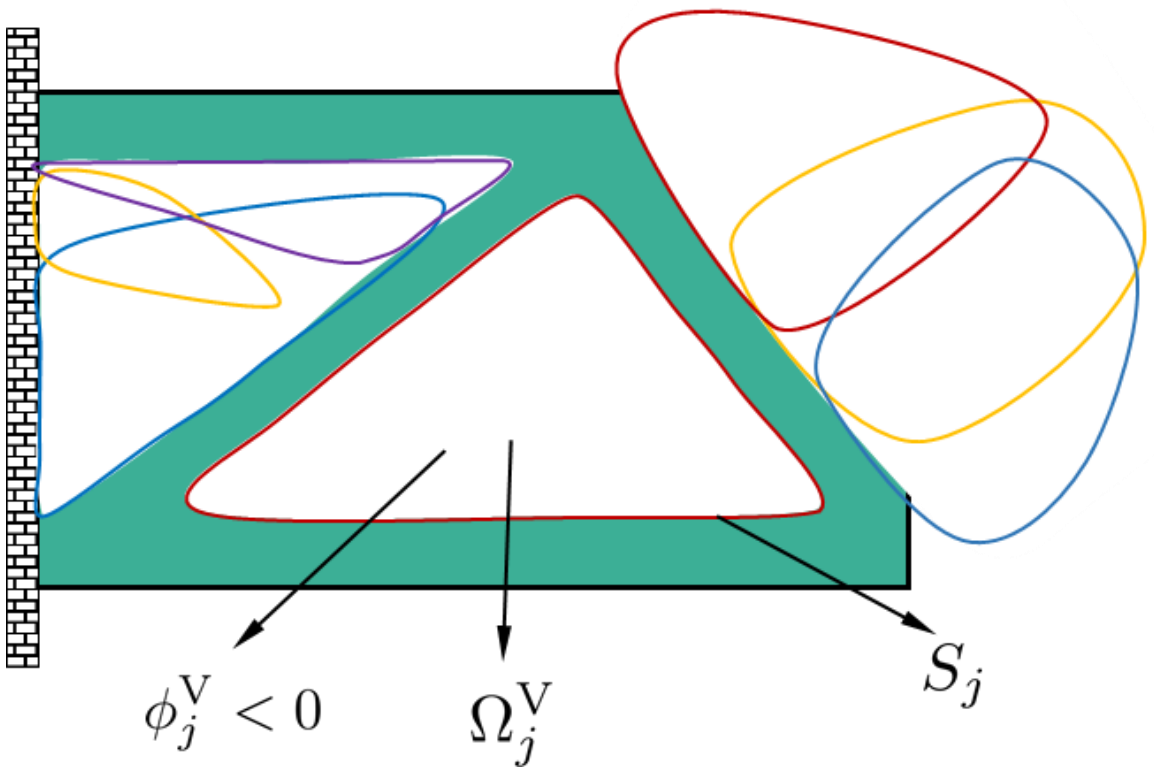

(c) The optimized layout of the voids.

Fig. 5 A schematic illustration of the MMV-based topology optimization method.

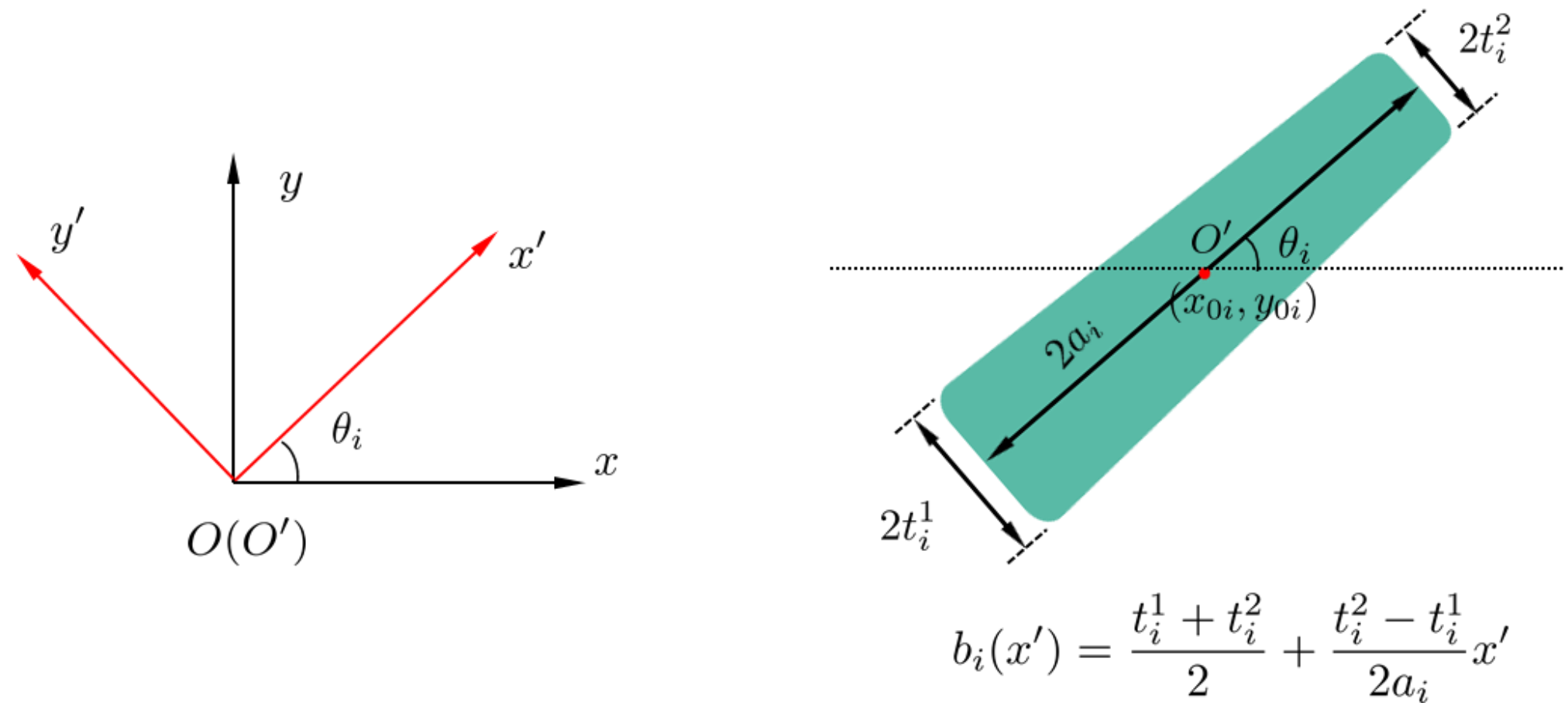

Fig. 6 The geometry description of a two-dimensional structural component.

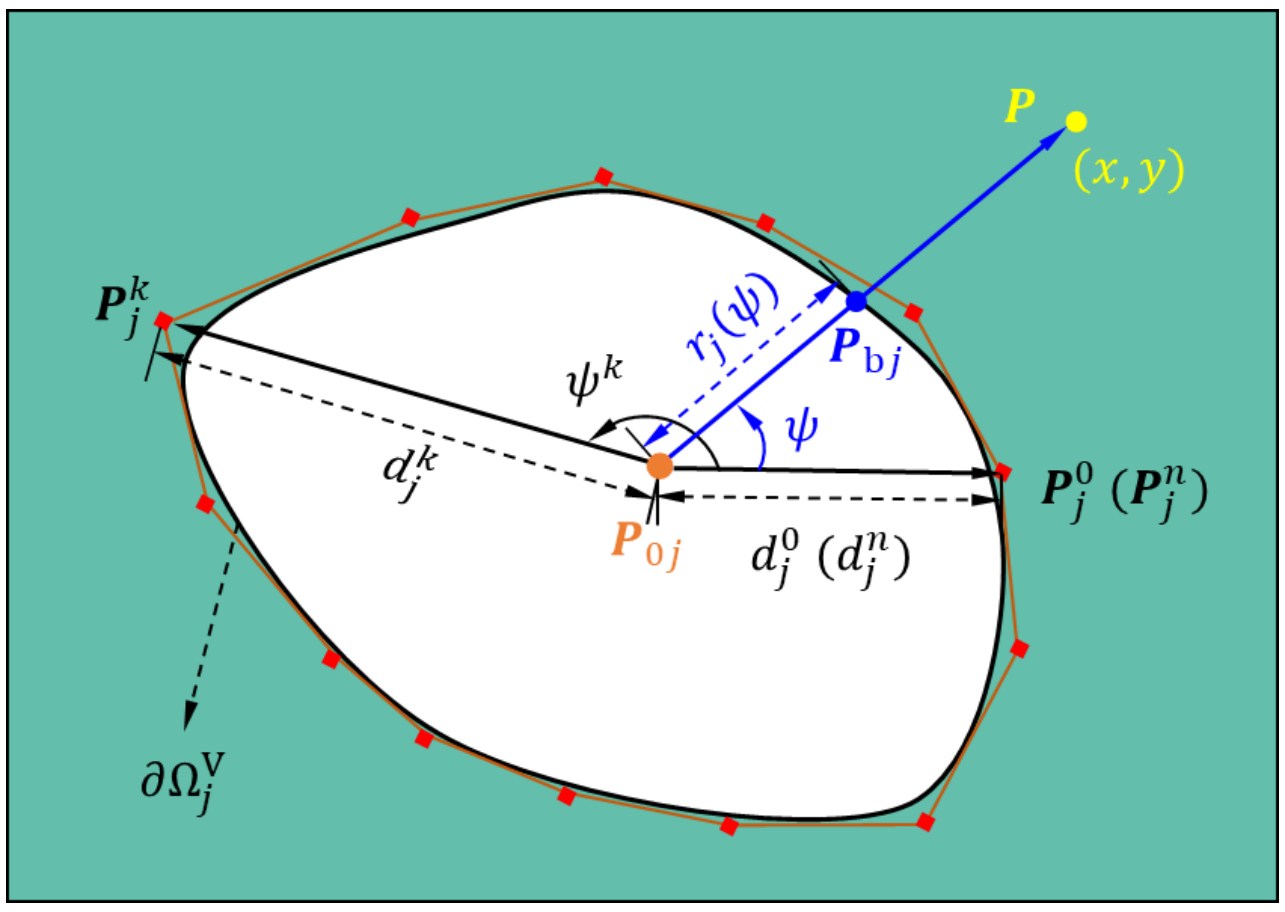

Fig. 7 The geometry description of a void in the MMV method.

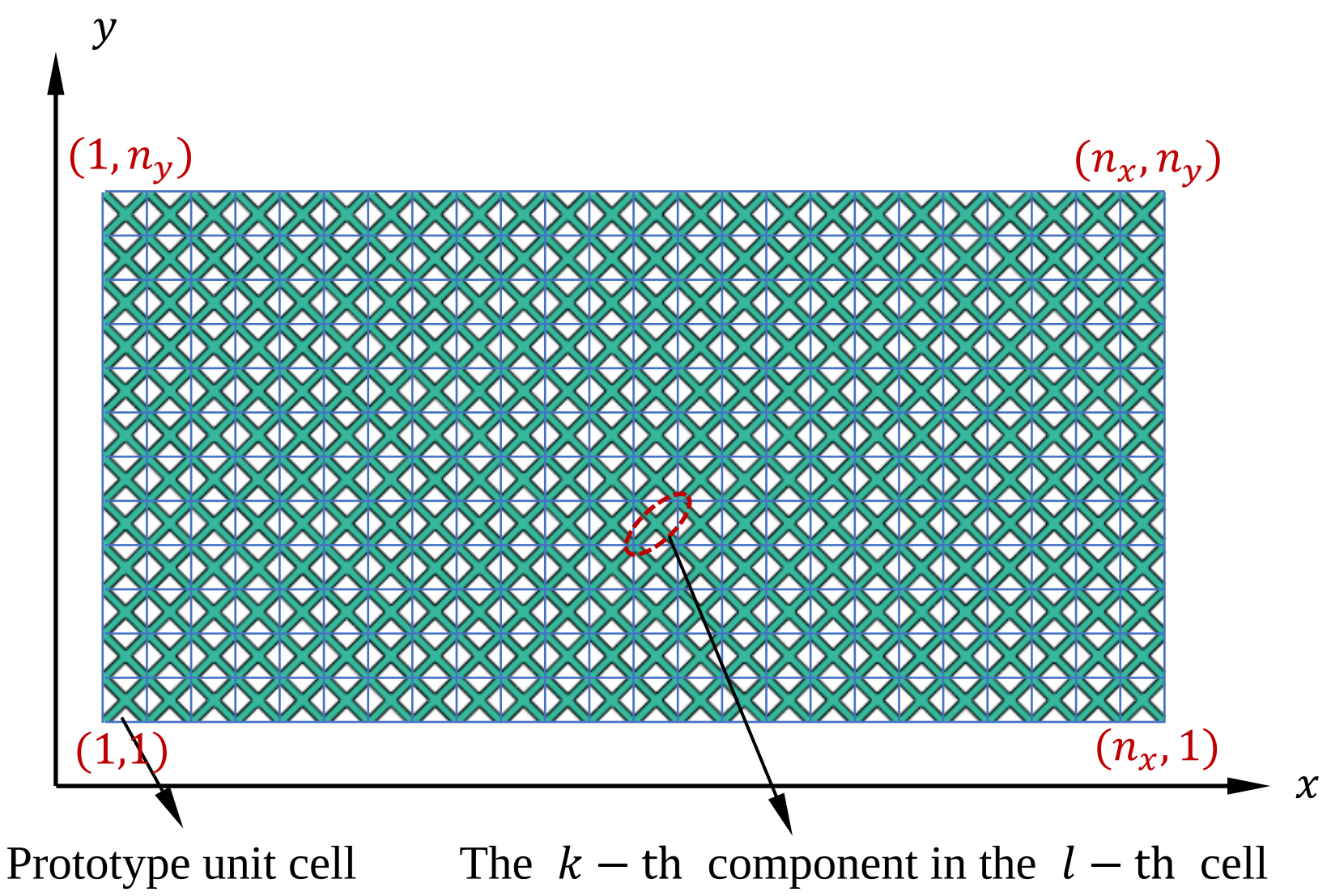

Fig. 8 Generating a periodic lattice structure through a prototype unit cell.

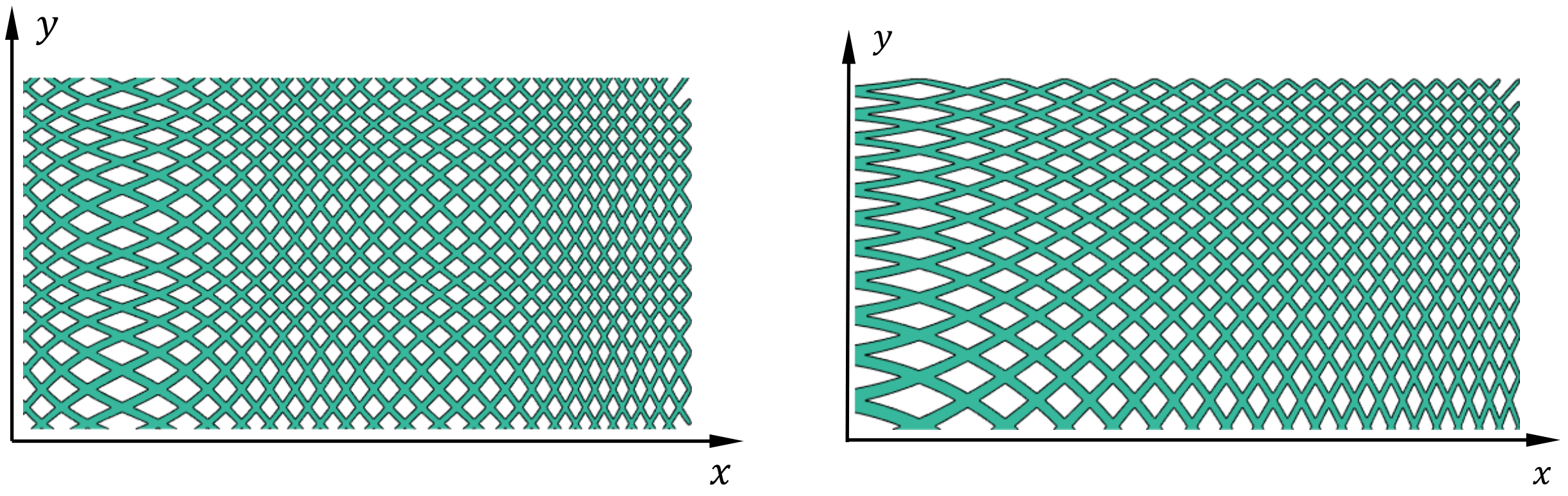

(a) Two trigonometric basis functions for each coordinate.

(b) Two polynomial basis functions for each coordinate.

Fig. 9 Generating a graded lattice structure through coordinate transformation of a periodic lattice in Fig. 8.

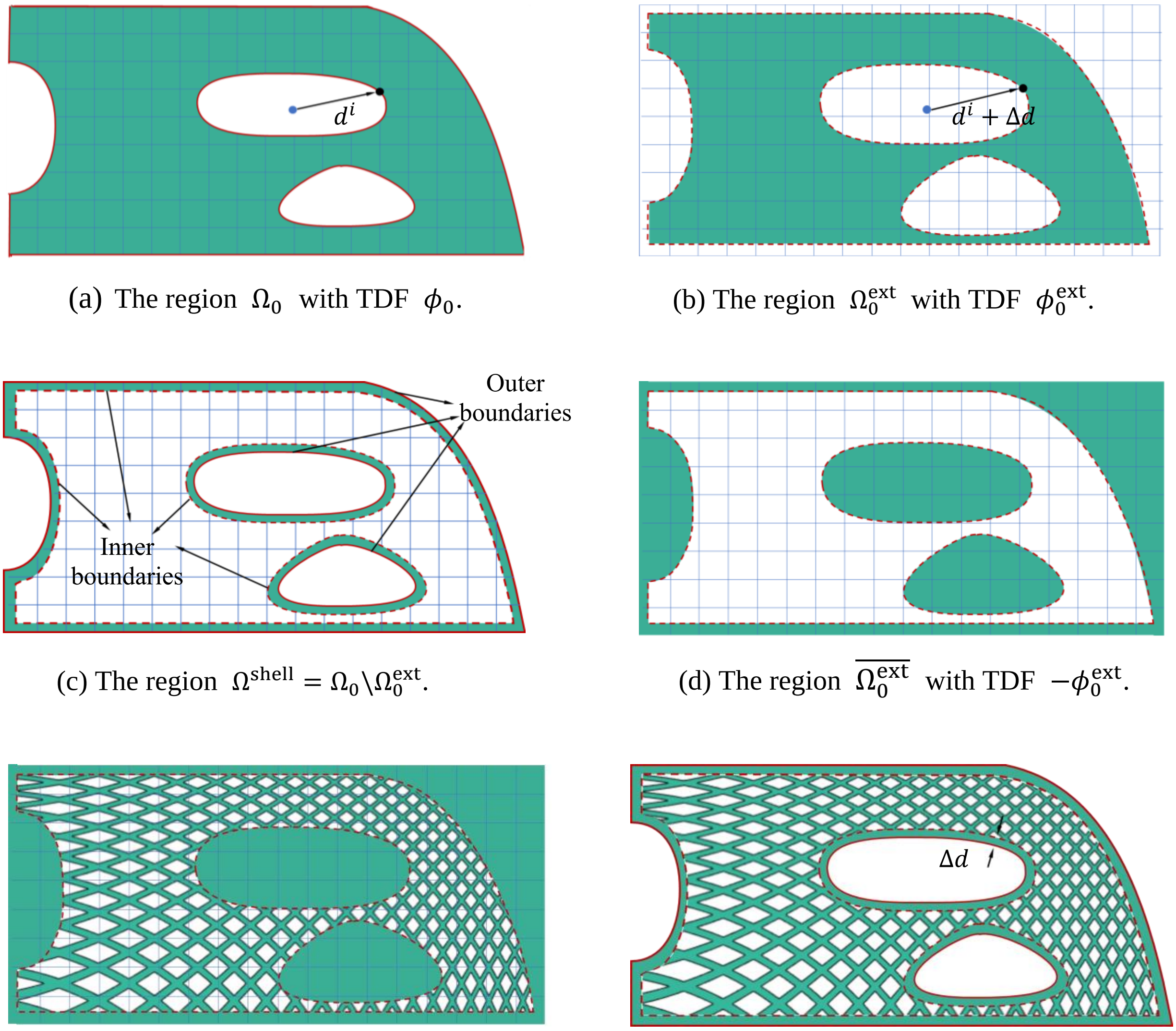

(a) The region $\Omega_0$ with TDF $\phi_0$.

(b) The region $\Omega_0^{\text{ext}}$ with TDF $\phi_0^{\text{ext}}$.

(c) The region $\Omega^{\text{shell}} = \Omega_0 \backslash \Omega_0^{\text{ext}}$.

(d) The region $\overline{\Omega_0^{\text{ext}}}$ with TDF $-\phi_0^{\text{ext}}$.

(e) The region $\Omega_1$ with TDF $\phi_1 = \max(-\phi_0^{\text{ext}}, \phi^{\text{g-s}})$.

(f) The region $\Omega^s$ with TDF $\phi^s = \min(\phi_0, \phi_1)$.

Fig.10 A schematic illustration of generating the TDF of a shell-graded-infill structure.

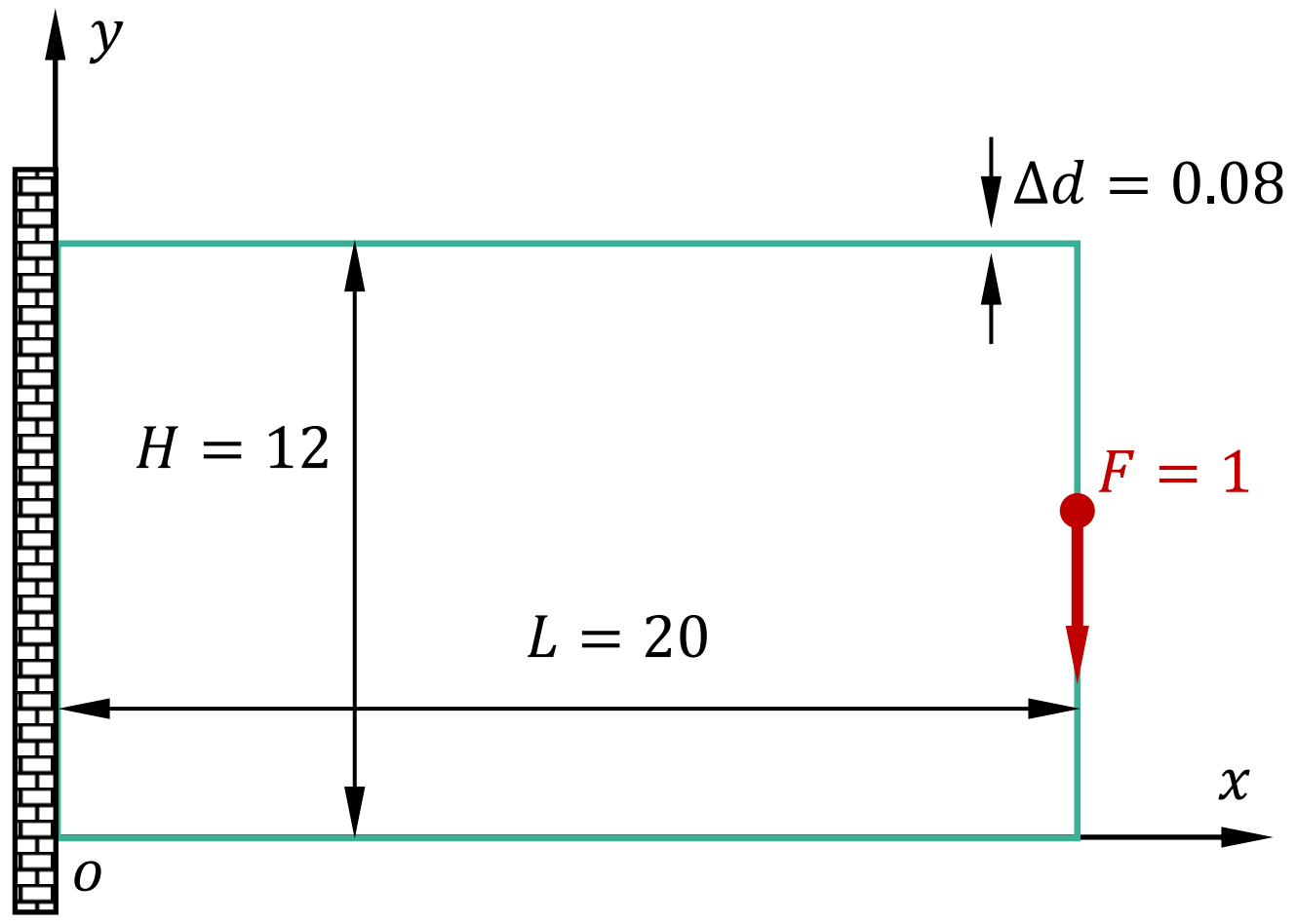

Fig. 11 The short beam example.

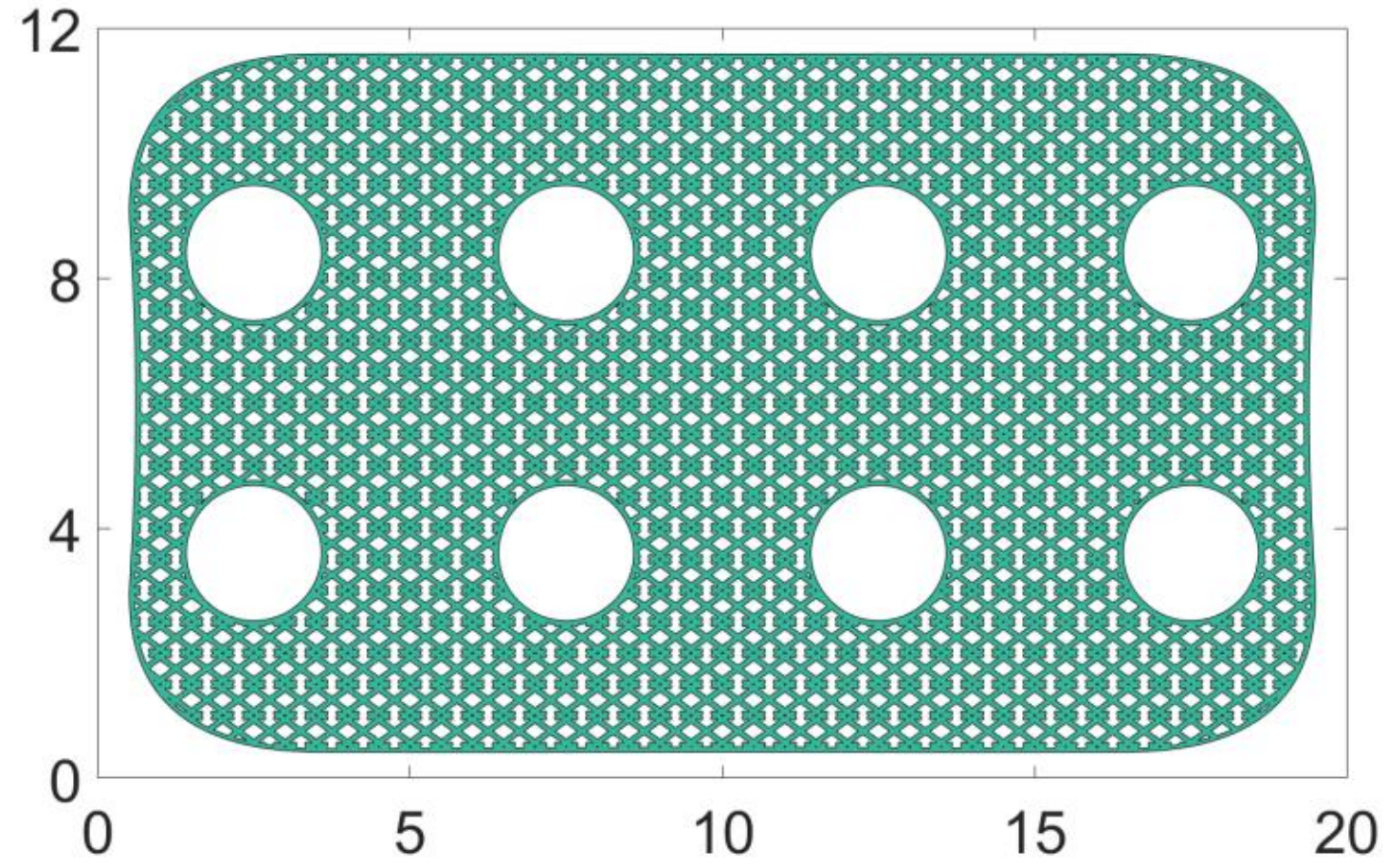

Fig. 12 The initial design of the short beam example.

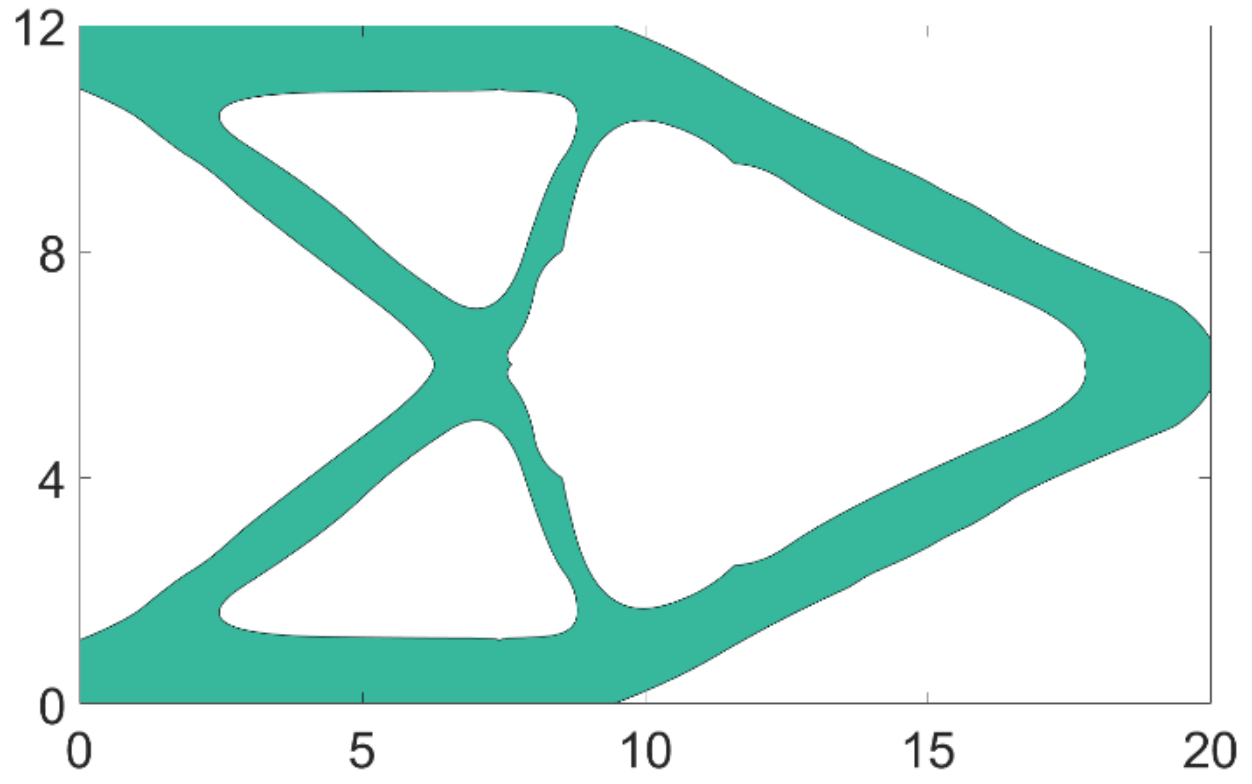

Fig. 13 The optimized topology of the short beam example without the volume constraint on the infill region ($C^{\mathrm{opt}} = 70.30$).

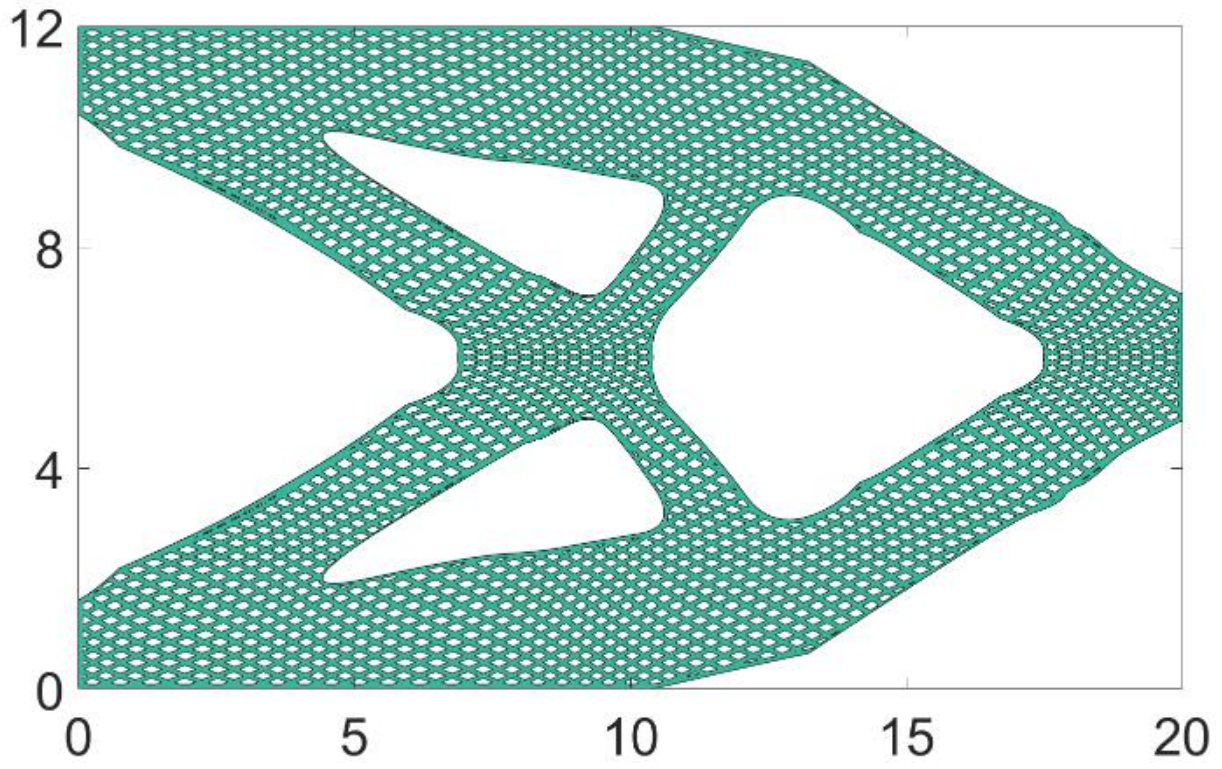

(a) $\underline{V} = 0.5V_{\mathrm{D}},\ C^{\mathrm{opt}} = 90.06.$

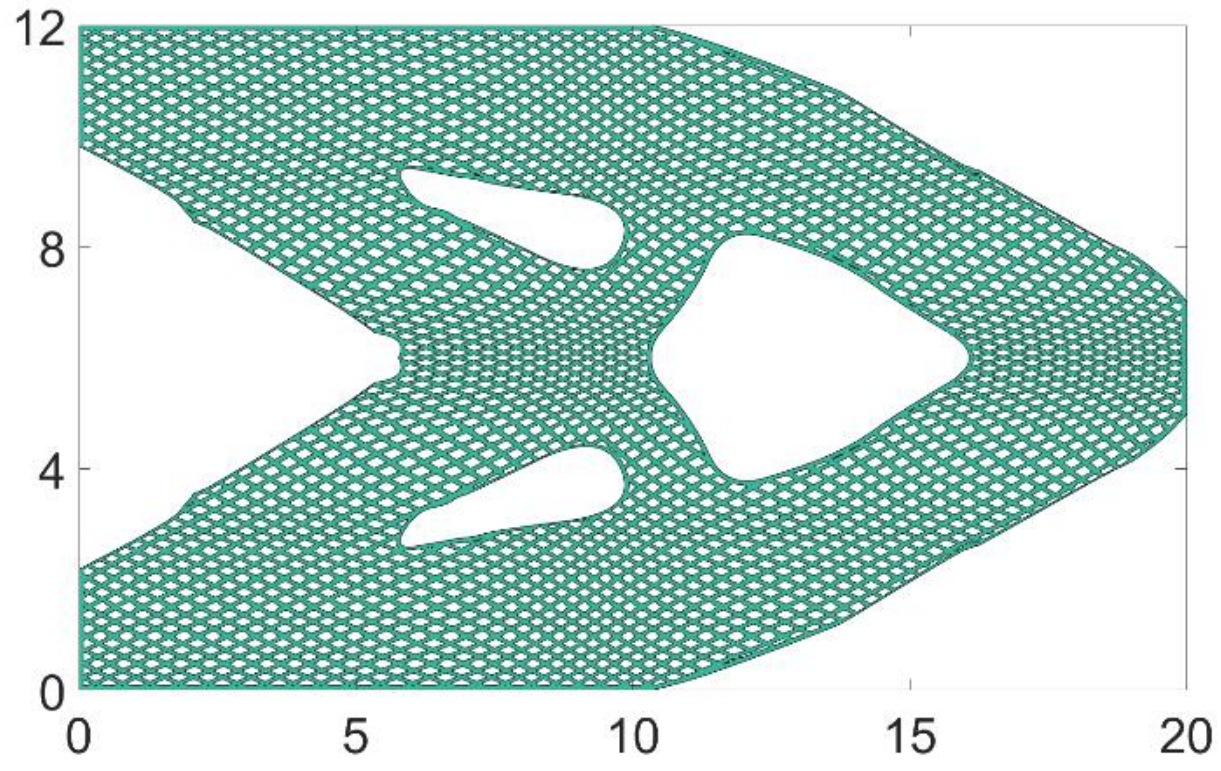

(b) $\underline{V} = 0.6V_{\mathrm{D}},\ C^{\mathrm{opt}} = 99.36.$

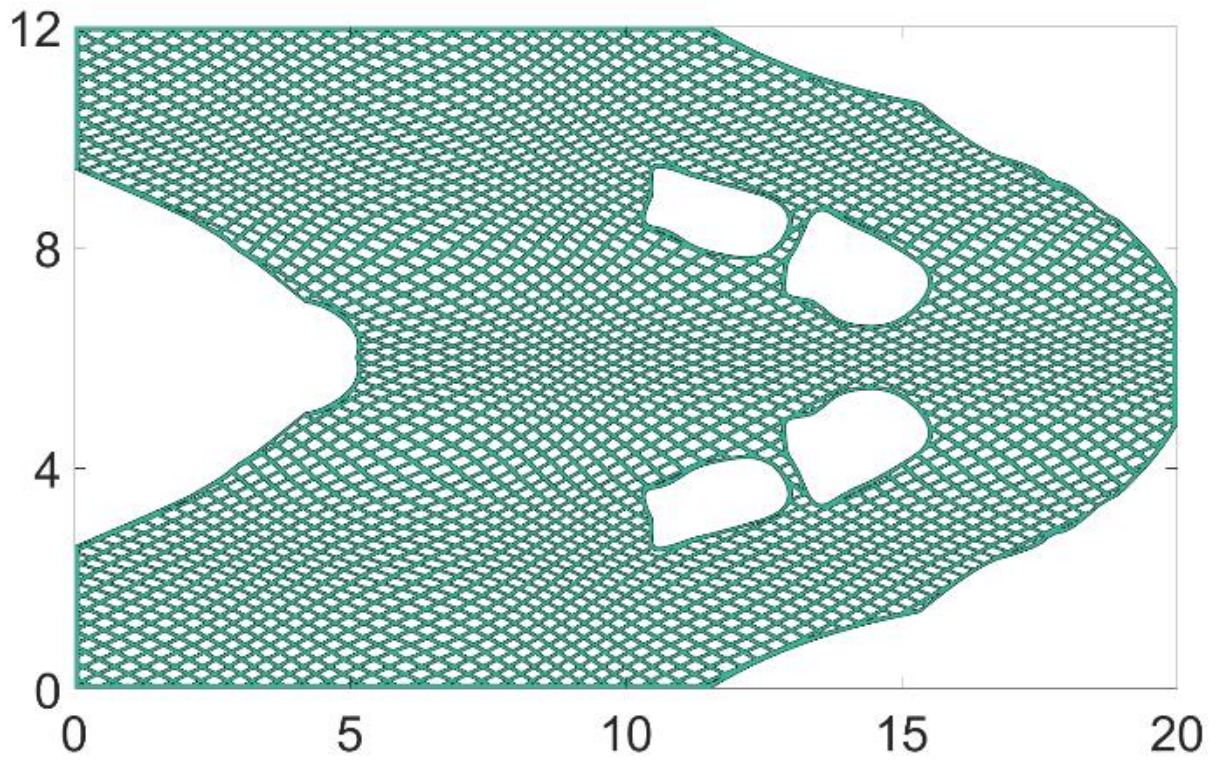

(c) $\underline{V} = 0.7V_{\mathrm{D}},\ C^{\mathrm{opt}} = 112.57.$

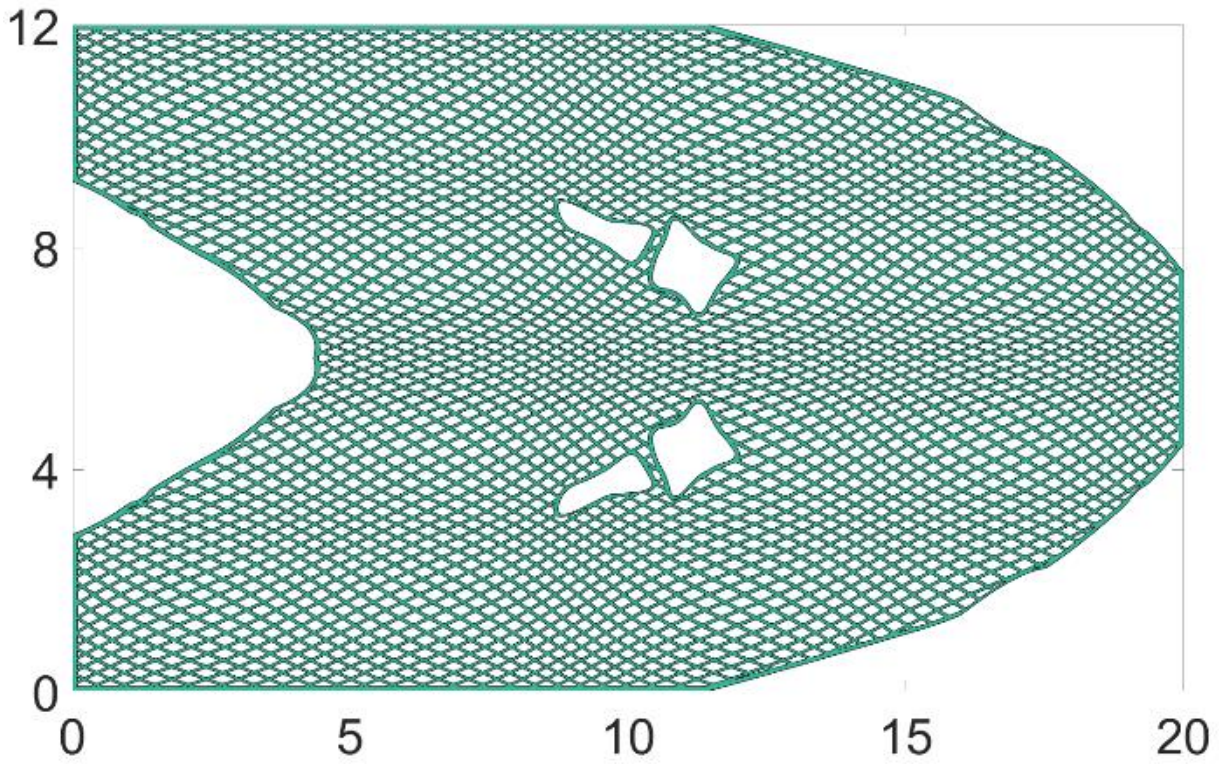

(d) $\underline{V} = 0.8V_{\mathrm{D}},\ C^{\mathrm{opt}} = 122.25.$

Fig. 14 The optimized results of the short beam example under different volume constraints on the infill region.

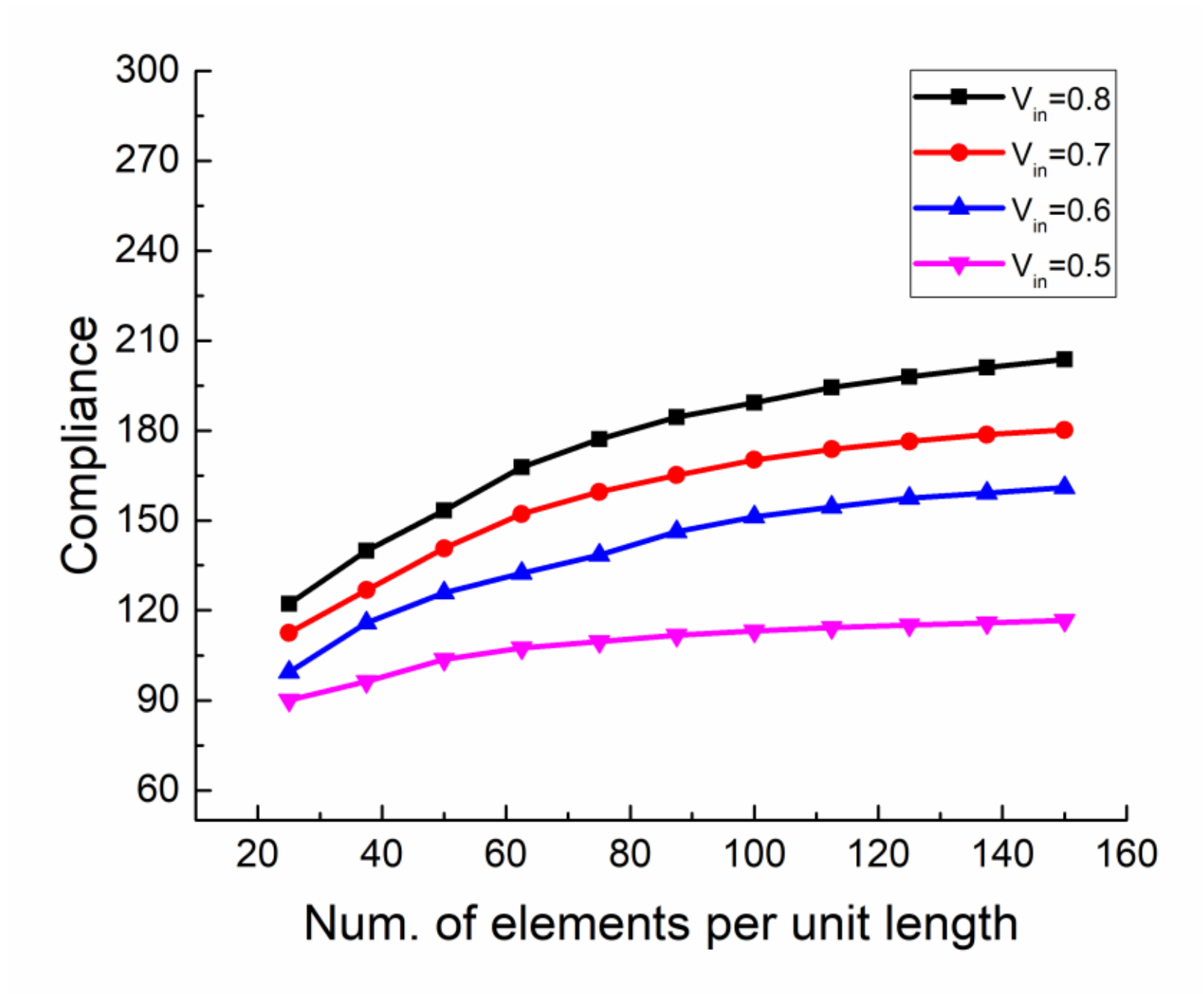

Fig. 15 The convergence test for the finite element analysis (short beam example).

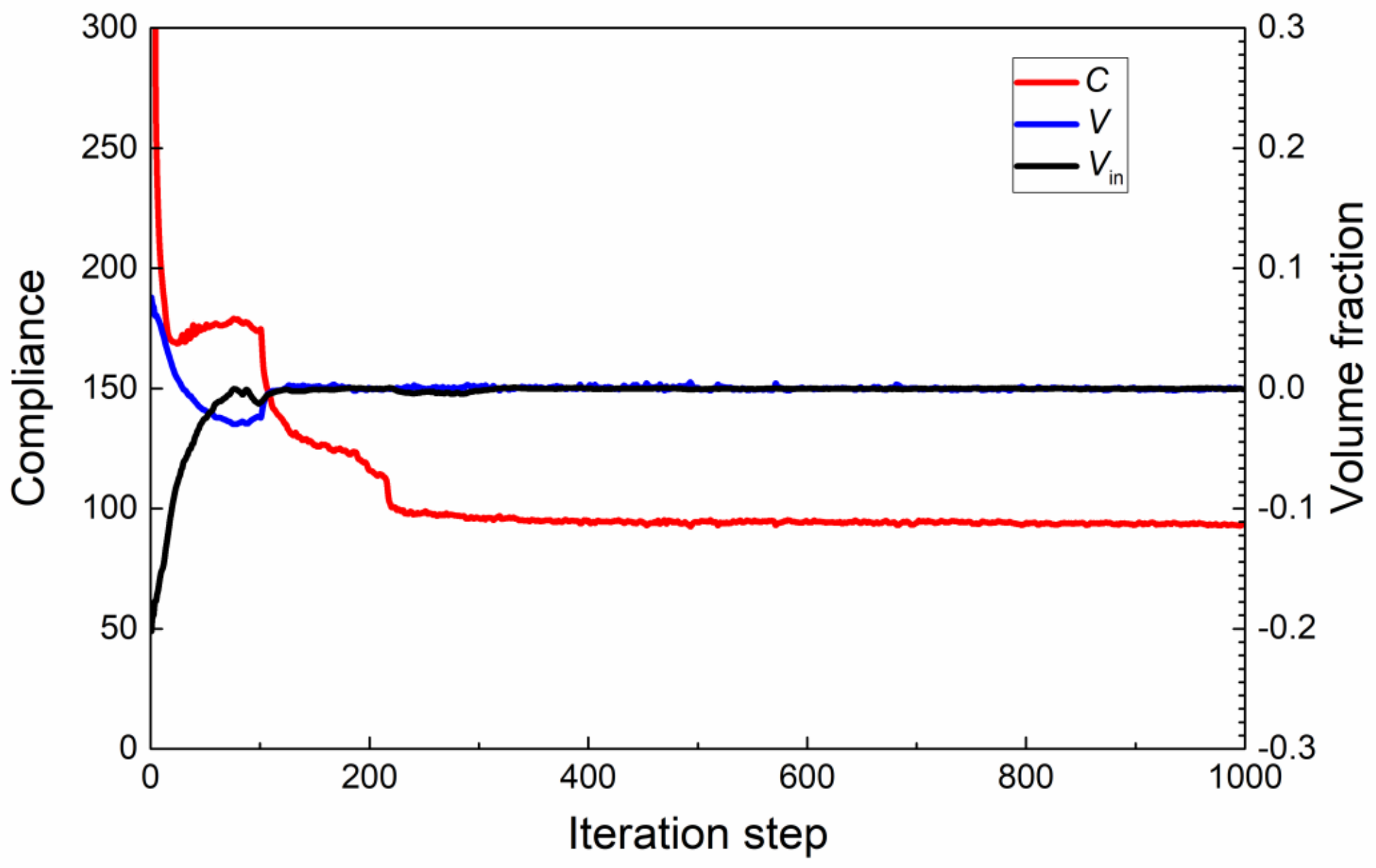

Fig. 16 The iteration history of the short beam example with $\underline{V} = 0.5V_{\mathrm{D}}$.

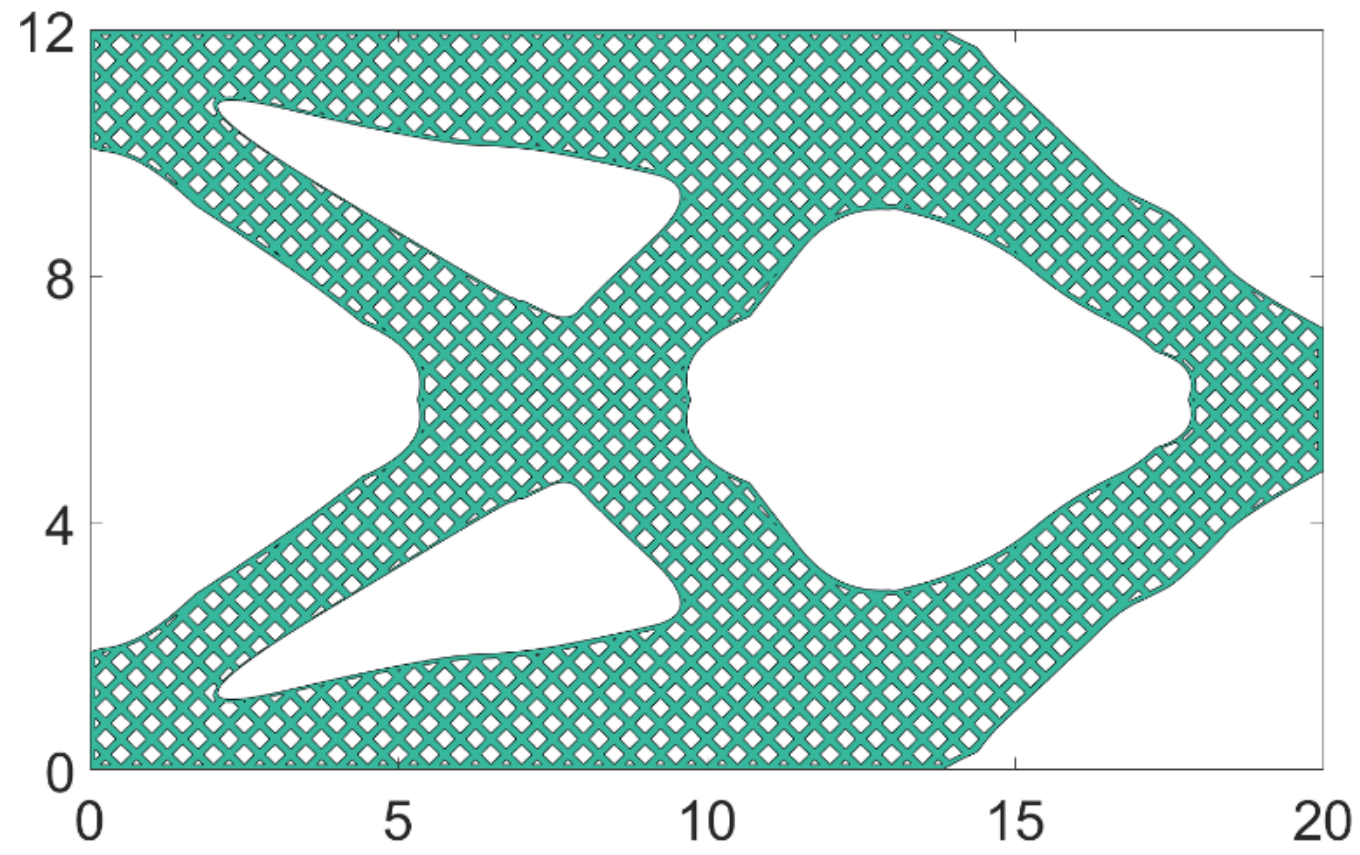

Fig. 17 The optimized coated structure with uniform infills

($\underline{V} = 0.5V_{\mathrm{D}}$ and $C^{\mathrm{opt}} = 134.15$).

.

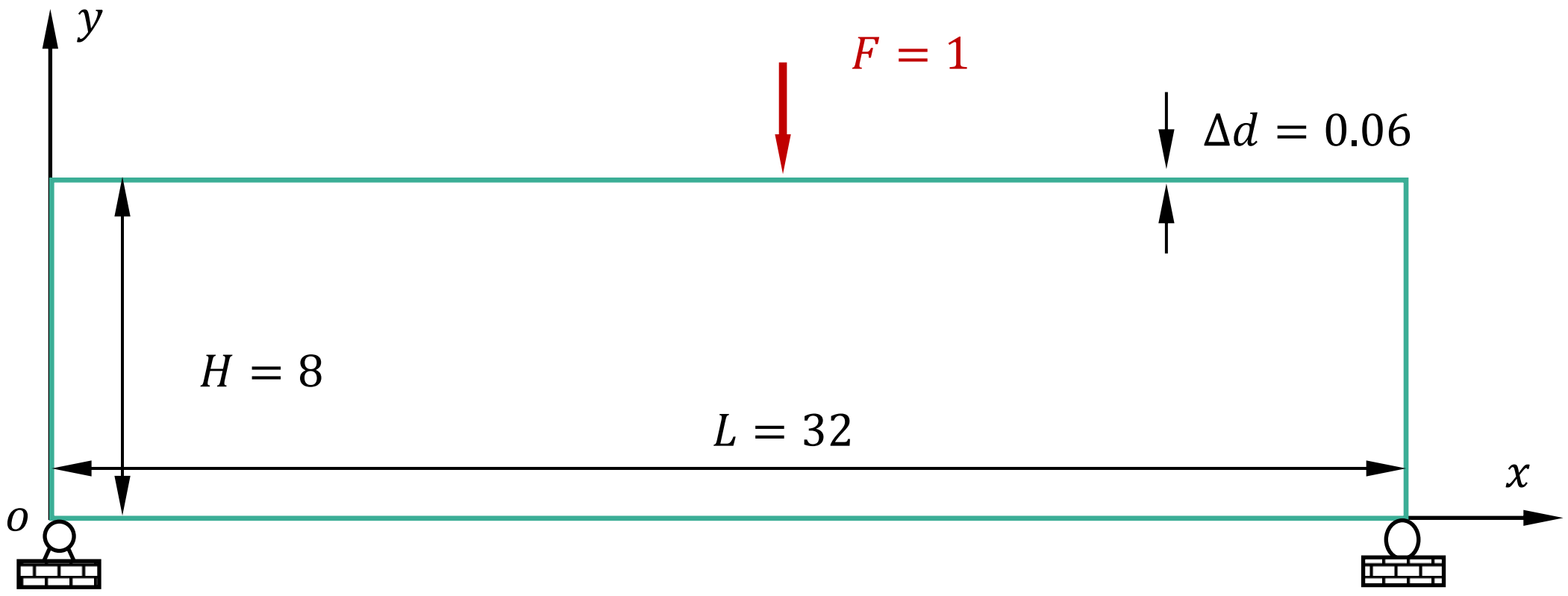

Fig. 18 The MBB beam example.

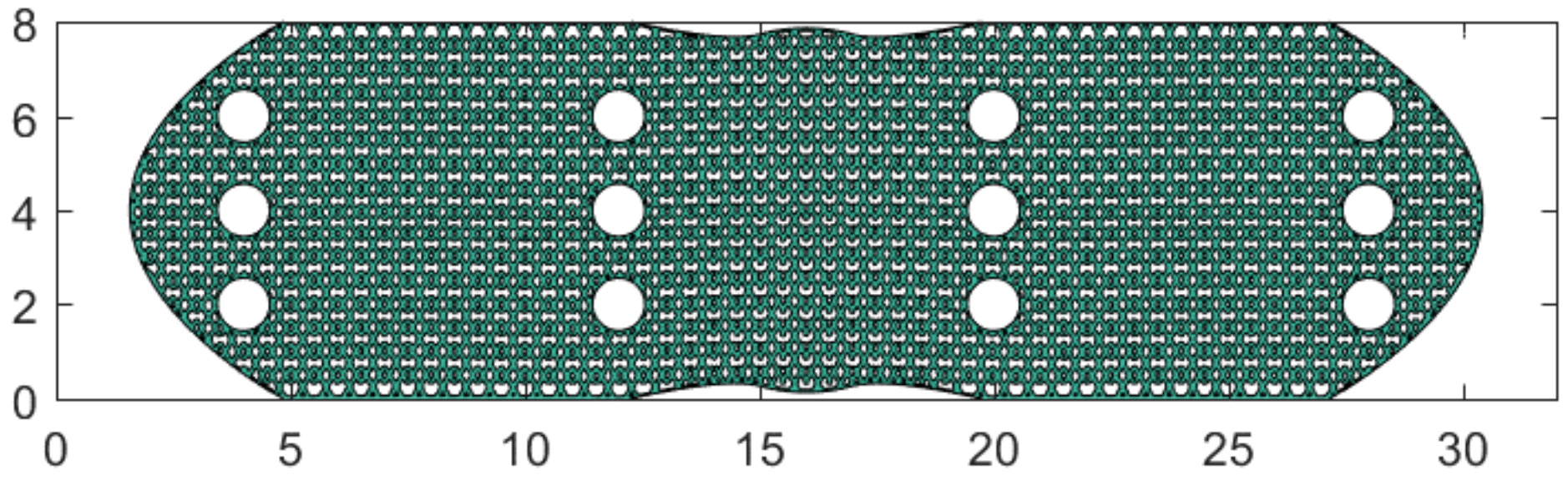

Fig. 19 The initial design of the MBB beam example.

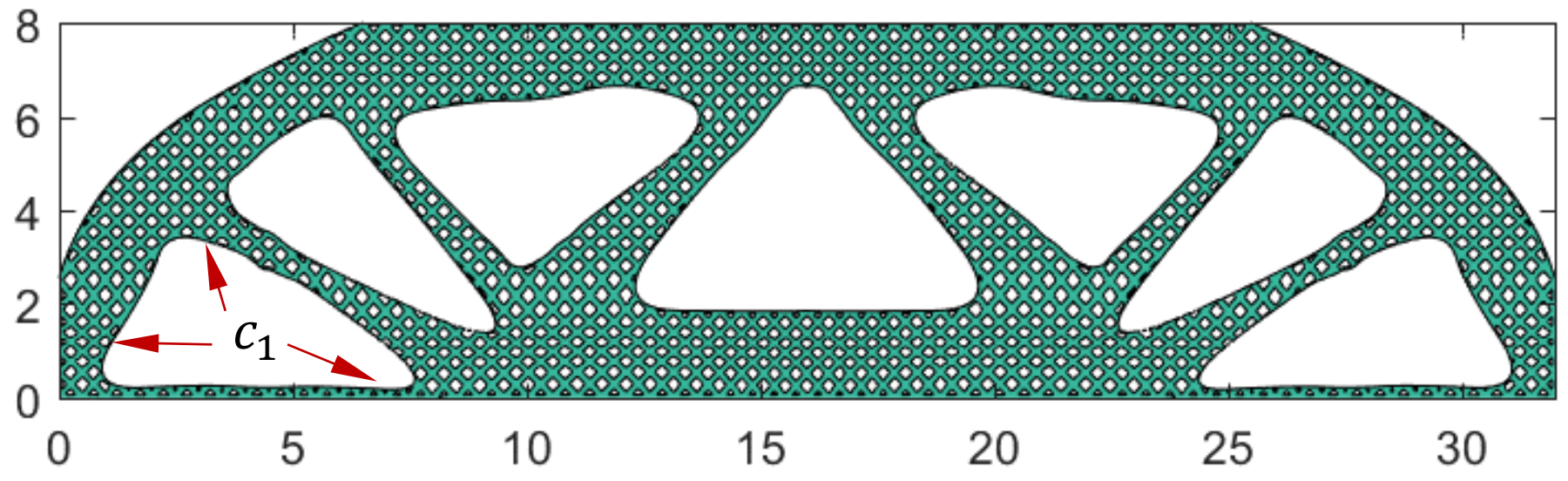

(a) $\underline{V} = 0.5V_{\mathrm{D}},\ C^{\mathrm{opt}} = 114.57.$

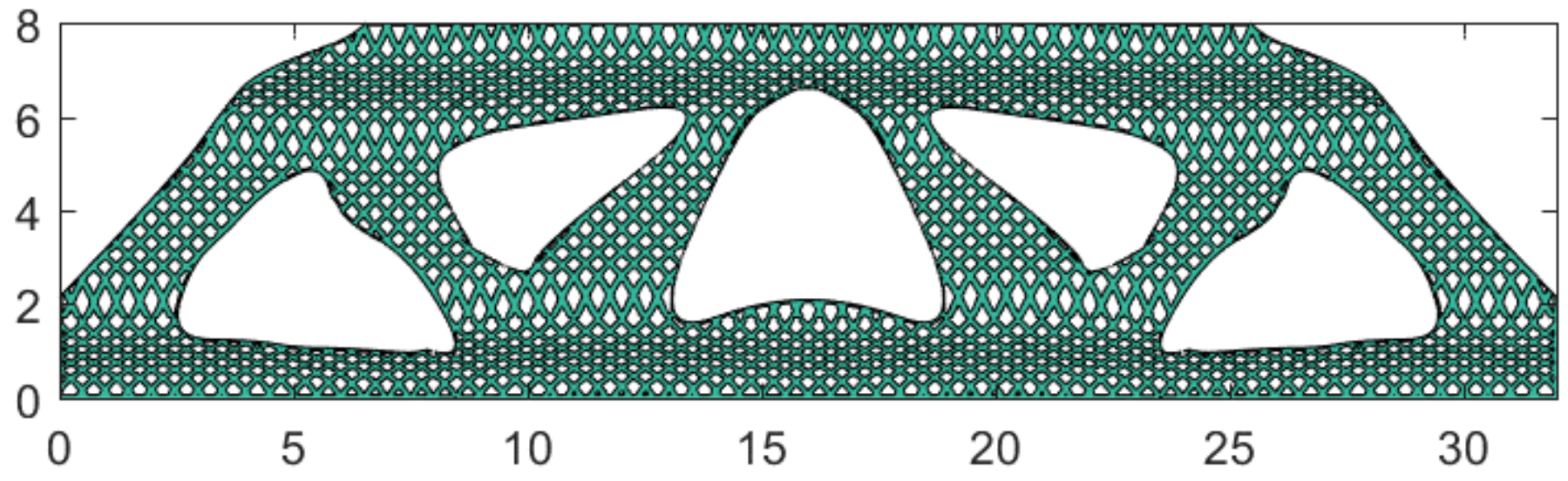

(b) $\underline{V} = 0.6V_{\mathrm{D}},\ C^{\mathrm{opt}} = 141.90.$

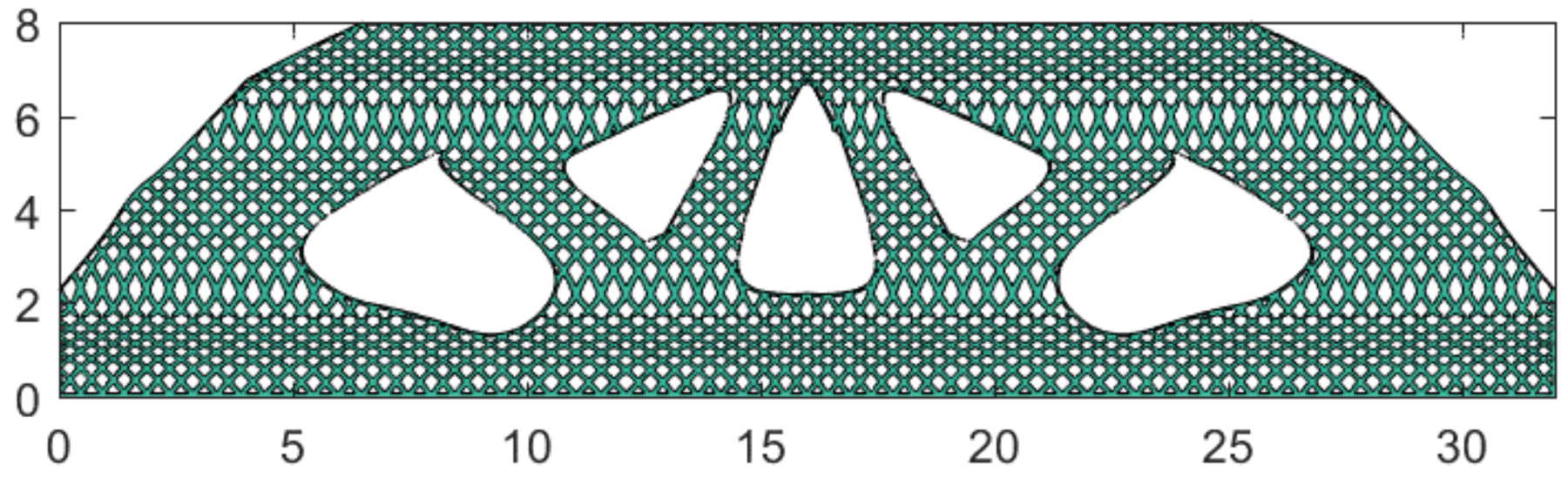

(c) $\underline{V} = 0.7V_{\mathrm{D}},\ C^{\mathrm{opt}} = 160.57.$

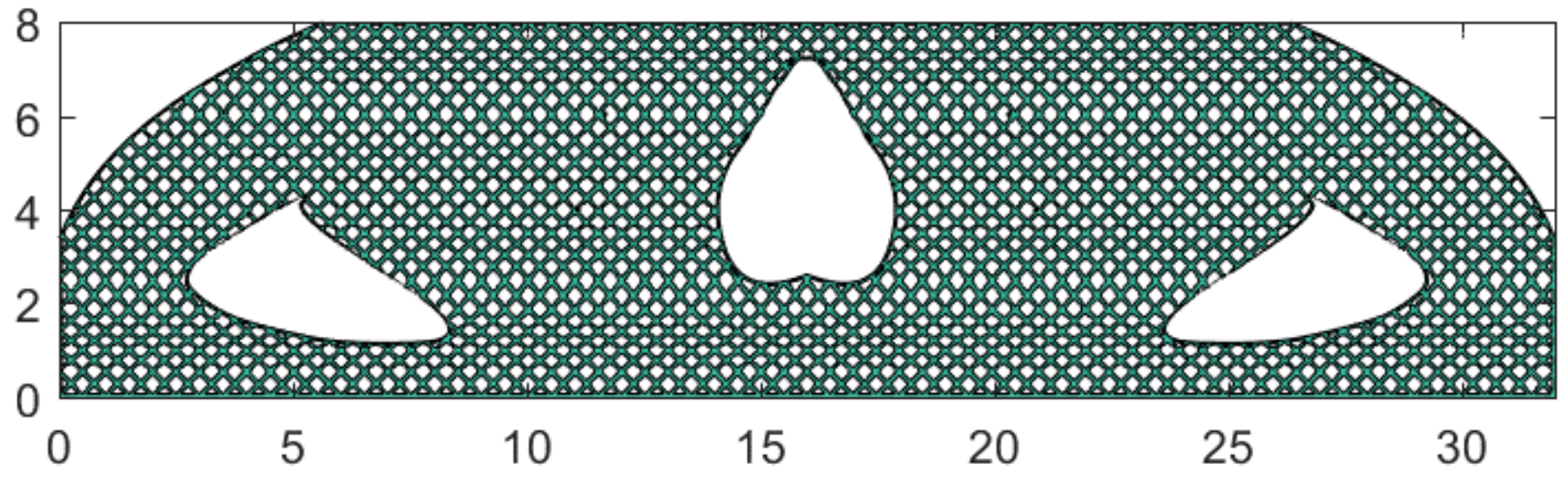

(d) $\underline{V} = 0.8V_{\mathrm{D}},\ C^{\mathrm{opt}} = 197.82.$

Fig. 20 The optimized results of the MBB beam example under different infill region volume constraints.

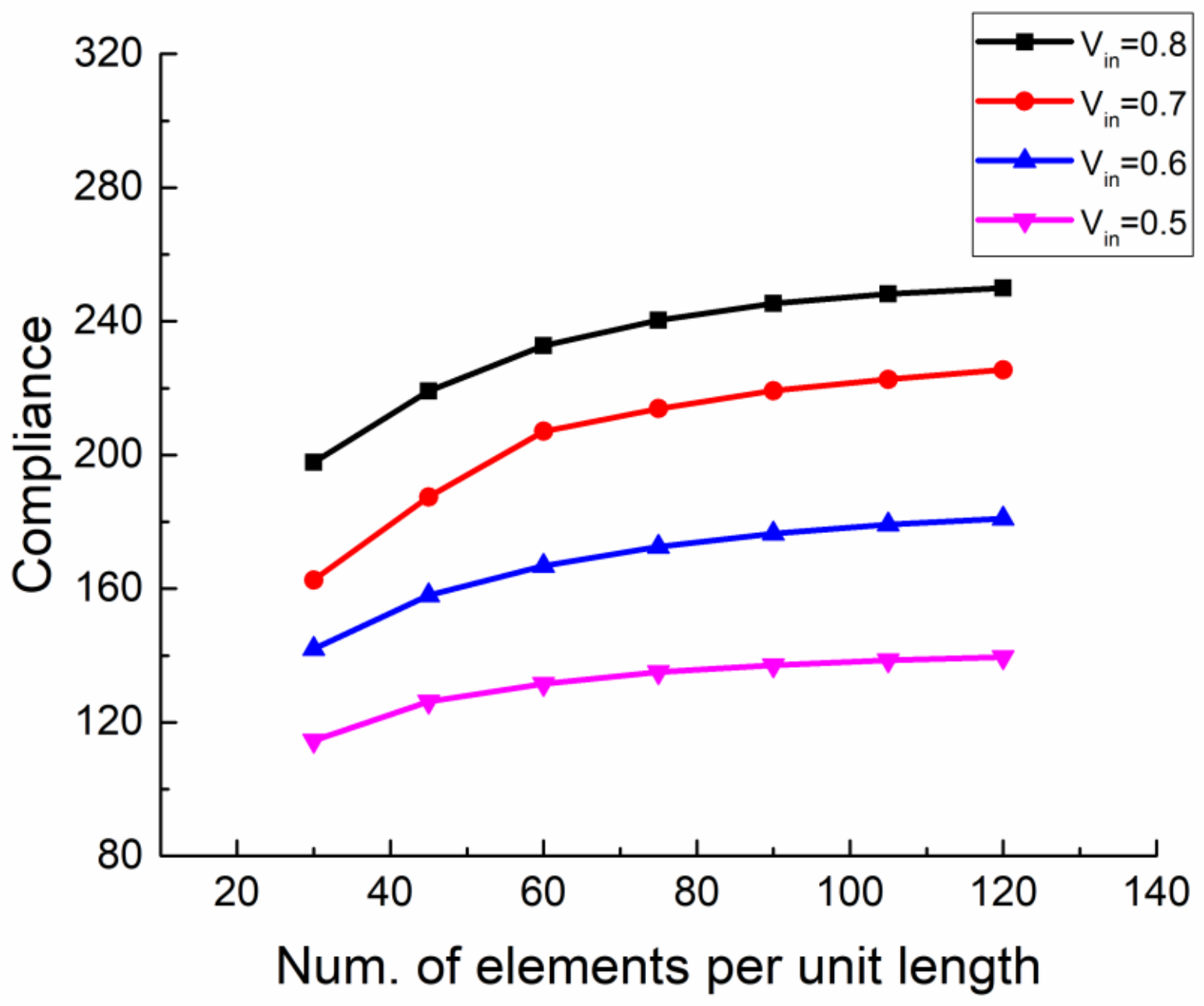

Fig. 21 Convergence test of finite element analysis of the MBB example.

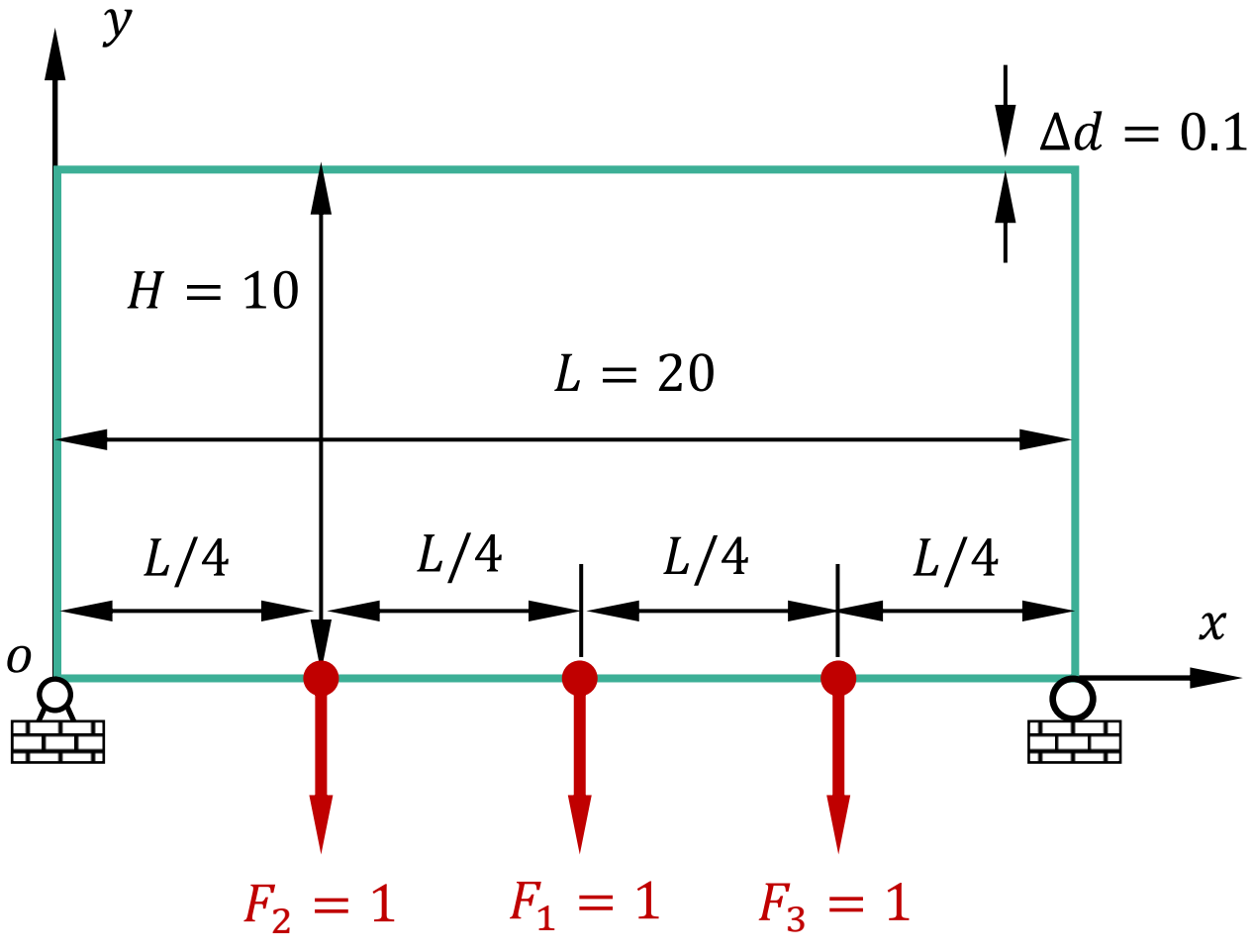

Fig. 22 The multi-load beam example.

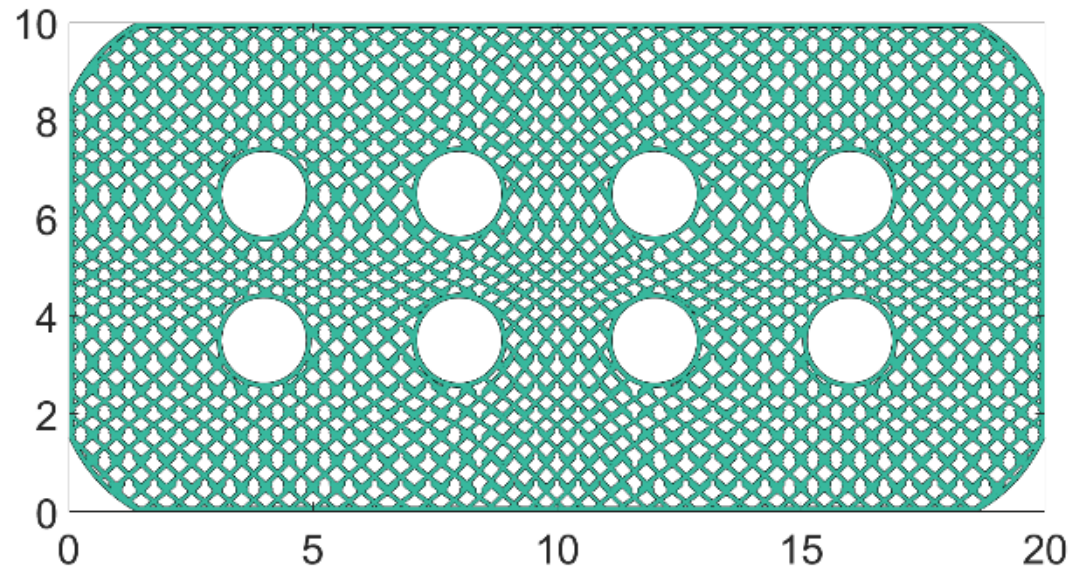

(a) 8 initial voids.

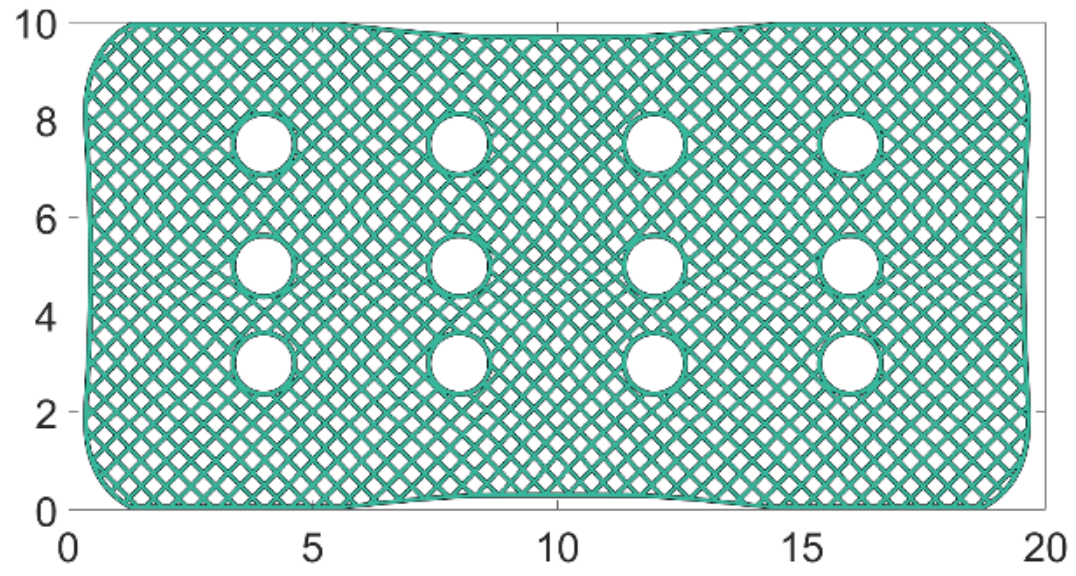

(b) 12 initial voids.

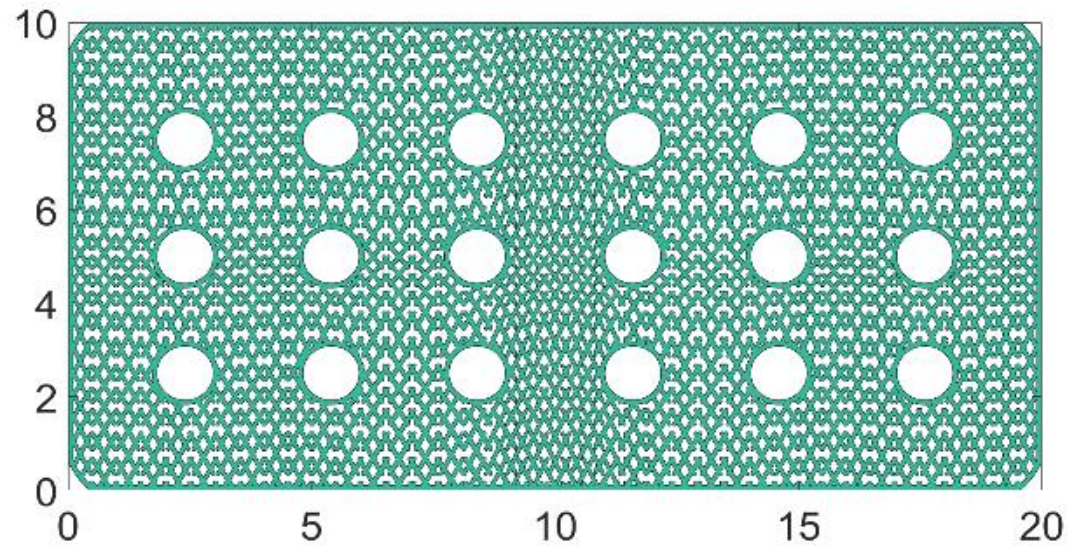

(c) 18 initial voids.

Fig. 23 The initial designs of the multi-load beam example.

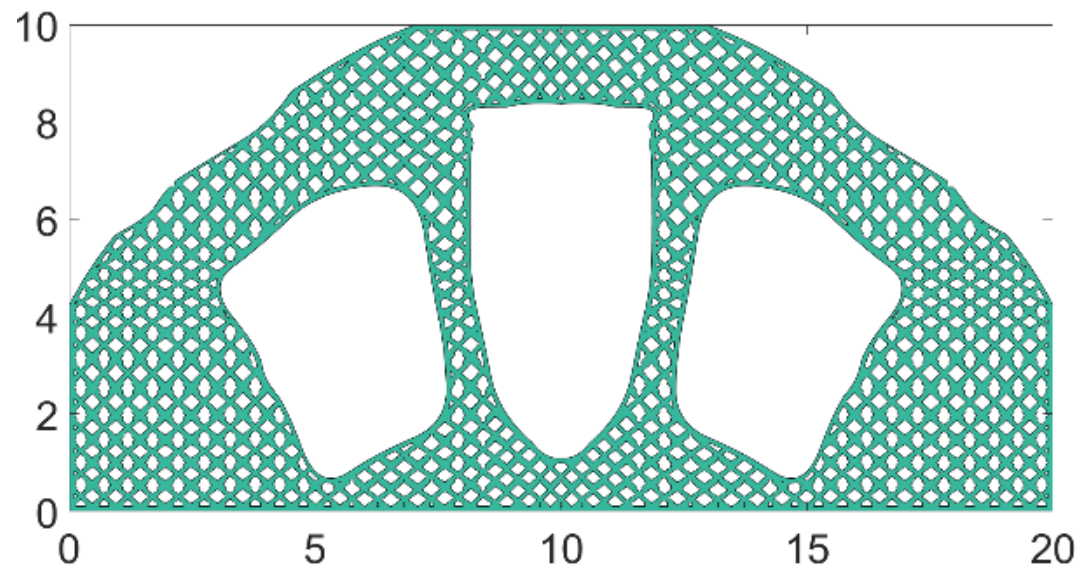

(a) 8 initial voids, ($C^{\text{opt}} = 243.67$).

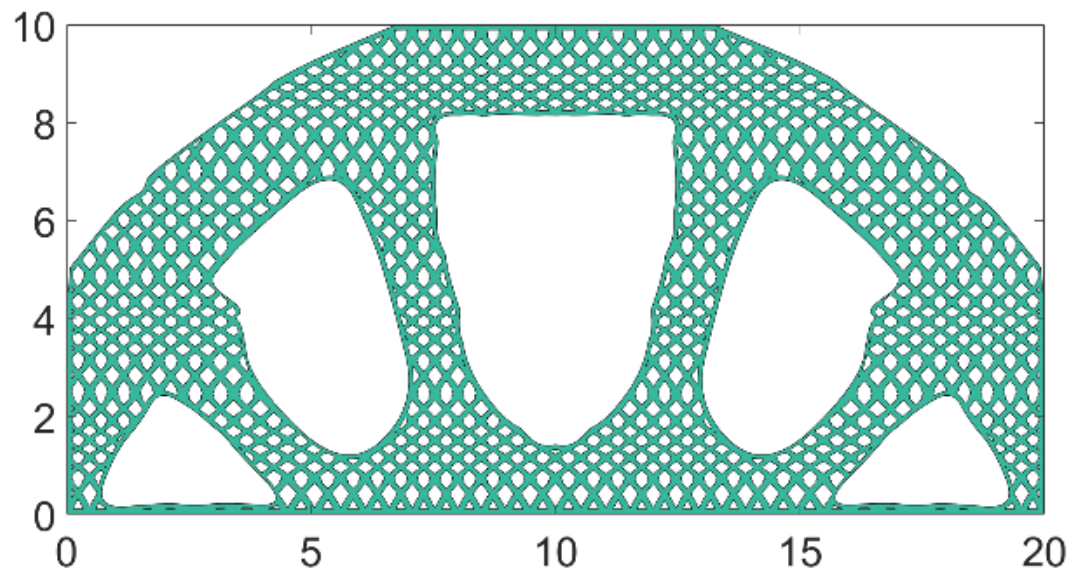

(b) 12 initial voids, ($C^{\text{opt}} = 232.68$).

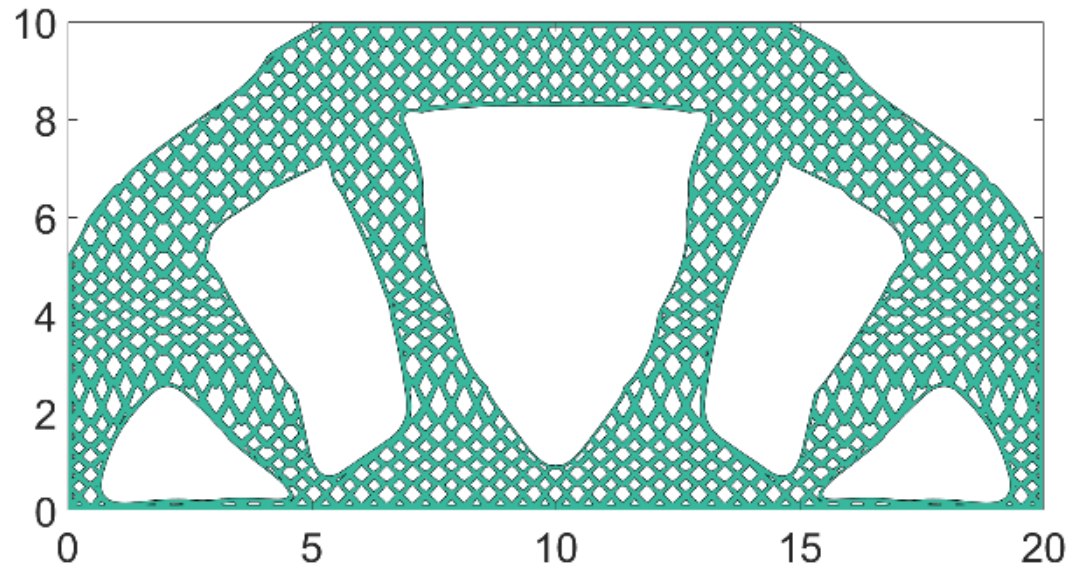

(c) 18 initial voids, ($C^{\text{opt}} = 232.03$).

Fig. 24 The optimized structures from different initial designs

(Single load case).

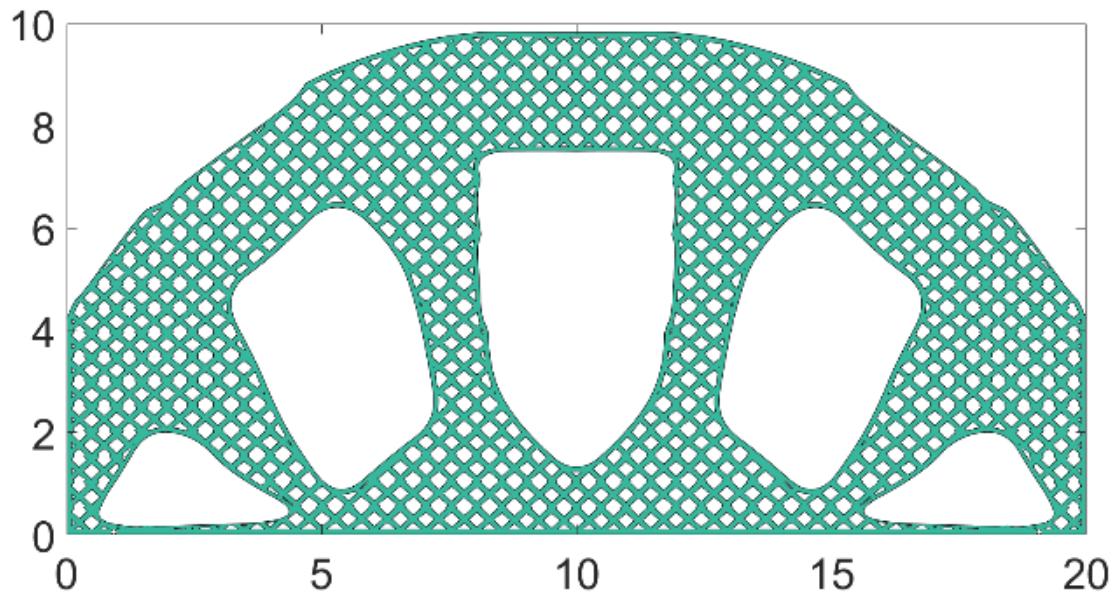

Fig. 25 The optimized structure obtained using two perturbation basis functions

(Single load case, $C^{\mathrm{opt}} = 243.08$).

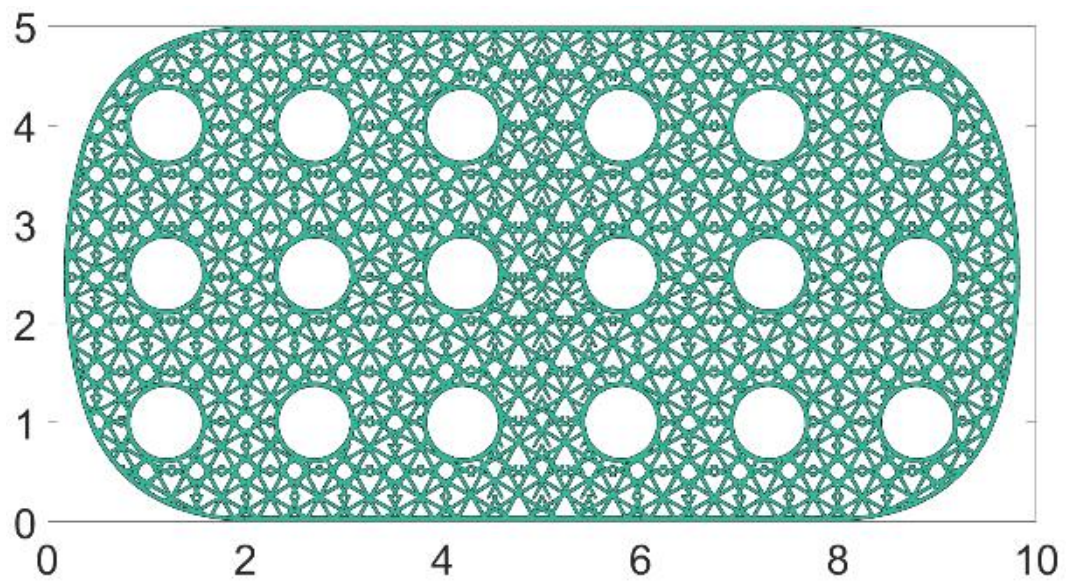

Fig. 26 The initial design with four components in the prototype unit cell.

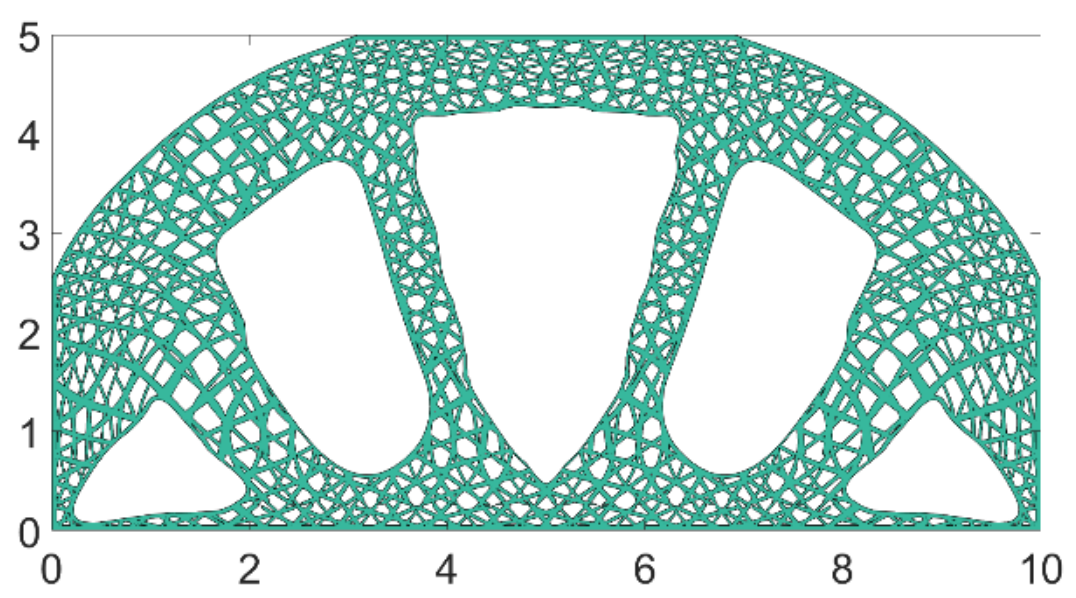

Fig. 27 The optimized structure obtained from the initial design where four components are locating in the prototype cell

(Single load case, $C^{\mathrm{opt}} = 213.28$).

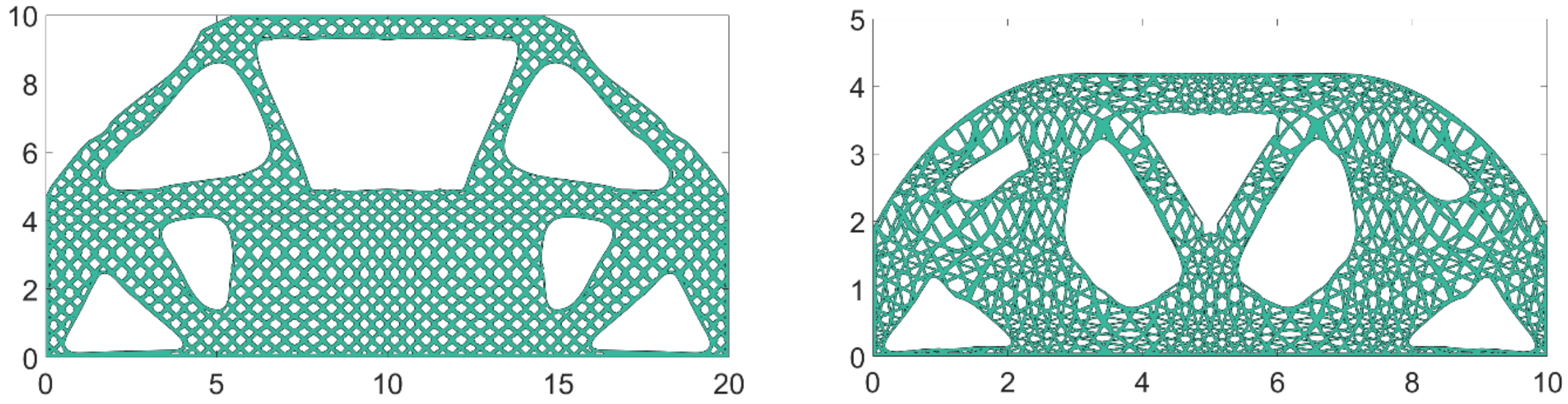

(a) 2 components in the primitive unit cell, ($C^{\mathrm{opt}} = 43.79$).

(b) 4 components in the primitive unit cell, ($C^{\mathrm{opt}} = 37.51$).

Fig. 28 The optimized structures obtained from the initial designs where different numbers of components are locating in the prototype unit cell (Multi-load case).